\documentstyle[12pt]{article}
\def\href#1#2{{#2}}
\begin{document}
\begin{titlepage}
\begin{flushright}
hep-ph/9501252 \\
THEP-95-1 \\
January 1995 \\
revised March 1995
\end{flushright}
\vspace{0.2cm}
\begin{center}
\LARGE
Gravity and the Standard Model \\
with $130$ $GeV$ Truth Quark \\
from $D_{4}-D_{5}-E_{6}$ Model \\
using $3 \times 3$ Octonion Matrices. \\
\vspace{1cm}
\normalsize
Frank D. (Tony) Smith, Jr. \\
\footnotesize
e-mail: gt0109e@prism.gatech.edu \\
and fsmith@pinet.aip.org  \\
P. O. Box for snail-mail: \\
P. O. Box 430, Cartersville, Georgia 30120 USA \\
\href{http://www.gatech.edu/tsmith/home.html}{WWW
URL http://www.gatech.edu/tsmith/home.html} \\
\vspace{12pt}
School of Physics  \\
Georgia Institute of Technology \\
Atlanta, Georgia 30332 \\
\vspace{0.2cm}
\end{center}
\normalsize
\begin{abstract}
The $D_{4}-D_{5}-E_{6}$ model of gravity and the Standard Model \\
with a $130$ $GeV$ truth quark is constructed using $3 \times 3$
matrices of octonions.  The model has both continuum
and lattice versions.  The lattice version uses
HyperDiamond lattice structure.
\end{abstract}
\vspace{0.2cm}
\normalsize
\footnoterule
\noindent
\footnotesize
\copyright 1995 Frank D. (Tony) Smith, Jr., Atlanta, Georgia USA
\end{titlepage}
\newpage
\setcounter{footnote}{0}
\setcounter{equation}{0}

\tableofcontents

\newpage

\section*{Foreword.}

In the $D_{4}-D_{5}-E_{6}$ mode, all tree level force strengths,
particle masses, and K-M parameters can be calculated.
They can be found at
\newline
\href{http://www.gatech.edu/tsmith/home.html}{WWW
URL http://www.gatech.edu/tsmith/home.html} \cite{SMI6}.
\newline
\vspace{12pt}

Within 10 \% or so, they are consistent with currently accepted
experimental values except for the truth quark mass.
Acccording to the $D_{4}-D_{5}-E_{6}$ model, it should
be $130$ $GeV$ at tree level, whereas CDF at Fermilab has
interpreted their data
to show that the truth quark mass is about $174 GeV$.
\newline
\vspace{12pt}

Fermilab's announcement is at
\newline
\href{http://fnnews.fnal.gov/}{WWW URL http://fnnews.fnal.gov/}
\cite{FNA}.
\newline
\vspace{12pt}

My opinion is that the CDF interpretation is wrong, and that
currently available experimental data indicates a truth quark
mass in the range of $120-145$ $GeV$, which is consistent with
the $130$ $GeV$ tree level calculation of the $D_{4}-D_{5}-E_{6}$
model.
\newline
\vspace{12pt}

Details of my opinion about the truth quark mass are at
\newline
\href{http://www.gatech.edu/tsmith/TCZ.html}{WWW
URL http://www.gatech.edu/tsmith/TCZ.html} \cite{SMI6}.
\newline
\vspace{12pt}

The $D_{4}-D_{5}-E_{6}$ model uses 3 copies of the octonions
to represent the 8 first generation fermion particles,
the 8 first generation fermion antiparticles, and
the 8-dimensional (before dimensional reduction) spacetime.
\newline
\vspace{12pt}

This paper is an effort to describe the construction
of the $D_{4}-D_{5}-E_{6}$ model by starting with
elementary structures, and building the model up
from them so that the interrelationships of the
parts of the model may be more clearly understood.
\newline
\vspace{12pt}

The starting points I have chosen are the octonions,
and $3 \times 3$ matrices of octonions \cite{FLY}.
\newline
\vspace{12pt}

Since the properties of octonions are not well covered
in most textbooks, the first section of the paper is
devoted to a brief summary of some of the properties
used in this paper.
\newline
\vspace{12pt}

$3 \times 3$ matrices are chosen because matrices are
well covered in conventional math and physics texts,
and also
\newline
because $3 \times 3$ traceless octonion matrices can be
used to build up in a concrete way the Lie algebras,
Jordan algebras, and symmetric spaces that are needed
to construct the $D_{4}-D_{5}-E_{6}$ model.
\newline
\vspace{12pt}

In particular, $3 \times 3$ traceless octonion matrices can be
divided into hermitian and antihermitian parts to form a Jordan
algebra $J_{3}^{\bf O}o$ and (when combined with derivations)
a Lie algebra $F_{4}$.
\newline
\vspace{12pt}

Then $J_{3}^{\bf O}o$ and $F_{4}$ can be combined to form
the Lie algebra $E_{6}$.
\newline
\vspace{12pt}

The components of the matrix representation of $E_{6}$ can
then be used to construct the Lagrangian
of the $D_{4}-D_{5}-E_{6}$ model of Gravity plus
the Standard Model for first generation fermions.
\newline
\vspace{12pt}

By extension to the exceptional Lie algebras
$E_{7}$ and $E_{8}$, the model includes the second
and third generation fermions.
\newline
\vspace{12pt}

Since the $E$-series of Lie algebras contains only
$E_{6}$, $E_{7}$ and $E_{8}$, there are only three
generation of fermions in the model.
\newline
\vspace{12pt}

\newpage

To get an overview of what this paper does,
represent the $E_{6}$ Lie algebra by

\begin{equation}
{\bf R} \otimes
\left(
\left(
\begin{array}{ccc}
S^{7}_{1} & {\bf O}_{+} &  {\bf O}_{v} \\
& & \\
-{\bf O}_{+}^\dagger & S^{7}_{2} &  {\bf O}_{-} \\
& & \\
-{\bf O}_{v}^\dagger & -{\bf O}_{-}^\dagger & -S^{7}_{1}-S^{7}_{2}
\end{array}
\right)
\oplus G_{2} \oplus
\left(
\begin{array}{ccc}
a & {\bf O}_{+} &  {\bf O}_{v} \\
& & \\
{\bf O}_{+}^\dagger & b &  {\bf O}_{-} \\
& & \\
{\bf O}_{v}^\dagger & {\bf O}_{-}^\dagger & -a-b
\end{array}
\right)
\right)
\end{equation}

where the {\bf R} represents the real scalar field
of this representation of $E_{6}$; ${\bf O}$ is octonion;
$S^{7}$ represents the imaginary octonions;
$a, b, c$ are real numbers (octonion real axis); and
$G_{2}$ is the Lie algebra of derivations of ${\bf O}$.

\vspace{12pt}

Then, use the components of $E_{6}$ to construct
the 8-dimensional Lagrangian of the $D_{4}-D_{5}-E_{6}$ model:

\begin{equation}
\int_{V_{8}} F_{8} \wedge \star F_{8} + \partial_{8}^{2}
\overline{\Phi_{8}} \wedge \star \partial_{8}^{2} \Phi_{8} +
\overline{S_{8\pm}} \not \! \partial_{8} S_{8\pm}
\end{equation}

where $\star$ is the Hodge dual;
\newline
\vspace{12pt}

$\partial_{8}$ is the 8-dimensional covariant derivative,
\newline
$\not \!  \partial_{8}$ is the 8-dimensional Dirac operator,
and
\newline
\vspace{12pt}

$F_{8}$ is the 28-dimensional $Spin(8)$ curvature,
\newline
which come from the $Spin(0,8)$ gauge group subgroup
of $E_{6}$, represented here by

\begin{equation}
\left(
\begin{array}{cc}
S^{7}_{1} & 0  \\
&  \\
0 & S^{7}_{2}
\end{array}
\right)
\oplus G_{2}
\end{equation}

\vspace{12pt}

$\Phi_{8}$ is the 8-dimensional scalar field,
which comes from the ${\bf R}$ scalar part of the
representation of $E_{6}$;
\newline
\vspace{12pt}

$V_{8}$ is 8-dimensional spacetime, and
which comes from the ${\bf O}_{v}$ part of the
representation of $E_{6}$;
\newline
\vspace{12pt}

$S_{8\pm}$ are the $+$ and $-$ 8-dimensional
half-spinor fermion spaces,
which come from the ${\bf O}_{+}$ and ${\bf O}_{-}$
parts of the representation of $E_{6}$;
\newline
\vspace{12pt}

As a theory with an 8-dimensional spacetime,
the $D_{4}-D_{5}-E_{6}$ model is seen to be constructed
from the fundamental representations of the $D_{4}$
Lie algebra $Spin(0,8)$:
\vspace{12pt}

$\Phi_{8}$ comes from the trivial scalar representation;
\vspace{12pt}

$F_{8}$ comes from the 28-dimensional adjoint representation;
\vspace{12pt}

$V_{8}$ comes from the 8-dimensional vector representation; and
\vspace{12pt}

$S_{8+}$ and $S_{8-}$ come from the two 8-dimensional
half-spinor representations.
\vspace{12pt}

As all representations of $Spin(0,8)$ can be built by
tensor products and sums using the 3 8-dimensional
representations and the 28-dimensional adjoint representation,
\newline
(the 4 of which make up the $D_{4}$ Dynkin diagram,
\newline
which looks like a Mercedes-Benz 3-pointed star,
\newline
with the 28-dimensional adjoint representation in the middle)
\newline
plus the trivial scalar representation,
\newline
the $D_{4}-D_{5}-E_{6}$ model effectively uses all the
ways you can look at $Spin(0,8)$.
\vspace{12pt}

If you use exterior wedge products as well as tensor
products and sums, you can build all the representations
from only the representations on the exterior of the
Dynkin diagram,
\newline
in this case, the 3 points of the star, the 3 8-dimensional
vector and half-spinor representations,
\newline
plus the trivial scalar representation.
\newline
In this case, it is clear because the adjoint representation
is the bivector representation, or the wedge product of
two copies of the vector representation.
\newline
For discussion of more complicated Lie algebras,
such as the $E_{8}$ Lie algebra, see Adams \cite{ADA2}.
\vspace{12pt}

Our physical spacetime is not 8-dimensional, and the
$D_{4}-D_{5}-E_{6}$ model gets to a 4-dimensional
spacetime by a process of dimensional reduction.
\vspace{12pt}

Dimensional reduction of spacetime to 4 dimensions
produces a realistic 4-dimensional Lagrangian of
Gravity plus the Standard Model.
\newline
\vspace{12pt}

Then force strength constants and particle masses are
calculated, as are Kobayashi-Maskawa parameters.
\newline
\vspace{12pt}

(The calculations are at tree level, with
quark masses being constituent masses.)
\newline
\vspace{12pt}

\vspace{12pt}

\vspace{12pt}

\vspace{12pt}

The structures in this paper are exceptional in many senses,
and can be studied from many points of view.

\vspace{12pt}

This paper is based on the structure of $3 \times 3$ matrices of
octonions.
\newline
For a discussion of $3 \times 3$ matrices of octonions from
a somewhat different perspective,
see Truini and Biedenharn \cite{TRU}.

\vspace{12pt}

I have also looked at the $D_{4}-D_{5}-E_{6}$ model
from the Clifford algebra point of view \cite{SMI3}.
\newline
Some useful references to Clifford algebras include
the books of Gilbert and Murray \cite{GILB},
of Harvey \cite{HAR}, and of Porteous \cite{POR}.

\vspace{12pt}

{}From the point of view of the exceptional Lie algebra $F_{4}$
and the Cayley Moufang plane ${\bf O}P^{2}$, see the paper
of Adams \cite{ADA1}.

\vspace{12pt}

{}From the point of view of the 7-sphere, the highest dimensional
sphere that is parallelizable, and the only parallelizable manifold
that is not a Lie group, see papers by
Cederwall and Preitschopf \cite{CED1},
Cederwall \cite{CED2},
Manogue and Schray \cite{MAN},
and
Schray and Manogue \cite{SCH}.

\vspace{12pt}

{}From the point of view of Hermitian Jordan Triple systems,
see papers by G\"{u}naydin \cite{GUN2,GUN3} and my paper \cite{SMI2}.

\vspace{12pt}

General references for differential geometry, Lie groups,
and the symmetric spaces used herein
are the books of Besse \cite{BES}, Fulton and Harris \cite{FUL},
Gilmore \cite{GIL},  Helgason \cite{HEL1}, Hua \cite{HUA},
Kobayashi and Nomizu \cite{KOB}, Porteous \cite{POR}, and
Postnikov \cite{POS}, and
the papers of Ramond \cite{RAM} and Sudbery \cite{SUD}.
\newline
An interesting discussion of symmetries in physics is
Saller \cite{SAL}.

\vspace{24pt}

A notational comment - this paper uses the same notation for a
Lie group as for its Lie algebra.  It should be clear from context
as to which is being discussed.

\vspace{24pt}

I would like to thank Geoffrey Dixon, Sarah Flynn,
David Finkelstein, Tang Zhong, Igor Kulikov, Michael Gibbs,
John Caputlu-Wilson, Ioannis Raptis, Julian Niles, Heinrich Saller,
Ernesto Rodriguez, Wolfgang Mantke, and Marc Kolodner
for very helpful discussions (some electronic) over the past year.

\newpage

\section{Octonions.}
There are 480 different ways to write a multiplication table
for an octonion product.  Since the purpose of this paper is to
construct a concrete representation of the $D_{4}-D_{5}-E_{6}$
physics model, one of these is chosen and used throughout.

\vspace{12pt}

For a good introduction to octonions, see the books of
Geoffrey Dixon \cite{DIX4} and of Jaak Lohmus, Eugene Paal, and
Leo Sorgsepp \cite{LOH}, as well as the paper of
G\"{u}naydin and G\"{u}rsey \cite{GUN1}.

\vspace{12pt}

For many more interesting things about octonions, see the book
and papers of Geoffrey Dixon
\cite{DIX1,DIX2,DIX3,DIX4,DIX5,DIX6,DIX7}.

\vspace{12pt}

The following description of octonion
products, left and right actions, and automorphisms and
derivations, is taken from Dixon's book and papers cited above,
and from a paper by A. Sudbery \cite{SUD}.


\subsection{Octonion Product.}

For concreteness, choose one of the 480 multiplications:
Let $e_{a}, a=1,...,7$, represent the imaginary units of
{\bf O}, and adopt the cyclic multiplication rule
\begin{equation}
e_{a}e_{a+1}=e_{a+5} = e_{a-2},
\end{equation}
a=1,...,7, all indices modulo 7, from 1 to 7
(another cyclic multiplication rule for {\bf O},
dual to that above, is $e_{a}e_{a+1}=e_{a+3} = e_{a-4}$).
In particular,
\begin{equation}
\{q_{1} \rightarrow e_{a}, q_{2} \rightarrow e_{a+1},
q_{3} \rightarrow e_{a+5}\}
\end{equation}
define injections of {\bf Q} into {\bf O} for a=1,...,7.
In the multiplication rule Equation (4) the indices range from 1 to 7,
and the index 0 representing the octonion real number $1$
is not subject to the rule.
\newline
This octonion multiplication has some very nice properties.
\vspace{12pt}

For example,
\begin{equation}
\mbox{if  } e_{a}e_{b}=e_{c}, \mbox{  then  } e_{(2a)}e_{(2b)}=e_{(2c)}.
\end{equation}
Equation (6) in combination with Equation (4) immediately implies
\begin{eqnarray}
& e_{a}e_{a+2}=e_{a+3}, \nonumber \\
& e_{a}e_{a+4}=e_{a+6}
\end{eqnarray}
(so  $e_{a}e_{a+2^{n}}=e_{a-2^{n+1}}$, or
$e_{a}e_{a+b} = [b^{3} \mbox{ mod } 7]e_{a-2b^{4}}$, $b=1,...,6$,
where $b^{3}$
out front provides the sign of the product (modulo 7,
$1^{3}=2^{3}=4^{3}=1$, and $3^{3}=5^{3}=6^{3}=-1$ )).
Also, 2(7)=7 mod 7, so Equations (6) and (4) imply
\begin{equation}
\begin{array}{ccc}
e_{7}e_{1}=e_{5}, & e_{7}e_{2}=e_{3}, & e_{7}e_{4}=e_{6}. \\
\end{array}
\end{equation}
These modulo 7 periodicity properties are reflected in the full
 multiplication table:
\begin{equation}
\left( \begin{array} {cccccccc}
1 & e_{1} & e_{2} & e_{3} & e_{4}  & e_{5} & e_{6} & e_{7}\\
e_{1}&-1&e_{6}&e_{4}&-e_{3}&e_{7}&-e_{2}&-e_{5}\\
e_{2}&-e_{6}&-1&e_{7}&e_{5}&-e_{4}&e_{1}&-e_{3}\\
e_{3}&-e_{4}&-e_{7}&-1&e_{1}&e_{6}&-e_{5}&e_{2}\\
e_{4}&e_{3}&-e_{5}&-e_{1}&-1&e_{2}&e_{7}&-e_{6}\\
e_{5}&-e_{7}&e_{4}&-e_{6}&-e_{2}&-1&e_{3}&e_{1}\\
e_{6}&e_{2}&-e_{1}&e_{5}&-e_{7}&-e_{3}&-1&e_{4}\\
e_{7}&e_{5}&e_{3}&-e_{2}&e_{6}&-e_{1}&-e_{4}&-1\\
\end{array} \right).
\end{equation}
Although the octonion product is nonassociative,
it is alternative.

\vspace{12pt}

An example of nonassociativity from the multiplication table is that
$$(e_{1}e_{2})e_{4} = e_{6}e_{2} = -e_{7}$$
is not equal to
$$e_{1}(e_{2}e_{4}) = e_{1}e_{5} = e_{7}$$

\vspace{12pt}

The alternativity property is the fact that the associator
\begin{equation}
[x,y,z] = x(yz) - (xy)z
\end{equation}
is an alternating function of $x,y,z \in \bf{O}$.

\subsection{Left and Right Adjoint Algebras of \bf O.}
The octonion algebra is nonassociative and so is not representable
as a matrix algebra.

\vspace{12pt}

However, the adjoint algebras of left and right actions of {\bf O}
on itself are associative.

\vspace{12pt}

For example, let    $u_{1},...,u_{n},x$ be elements of {\bf O}.
Consider the left adjoint map
\begin{equation}
x \rightarrow u_{n}(...(u_{2}(u_{1}x))...).
\end{equation}
The nesting of parentheses forces the products to occur in a
certain order, hence this algebra of left-actions is trivially
associative, and it is representable by a matrix algebra.

\vspace{12pt}

One such representation can be derived immediately from the
multiplication table Equation (9).  For example, the actions
\begin{equation}
x \rightarrow e_{1}x \equiv e_{L1}(x), \mbox{ and
} x \rightarrow xe_{1} \equiv e_{R1}(x)
\end{equation}
can be identified with the matrices
\begin{equation}
e_{L1} \rightarrow  \left( \begin{array} {cccccccc}
.&-1&.&.&.&.&.&.\\
1&.&.&.&.&.&.&.\\
.&.&.&.&.&.&-1&.\\
.&.&.&.&-1&.&.&.\\
.&.&.&1&.&.&.&.\\
.&.&.&.&.&.&.&-1\\
.&.&1&.&.&.&.&.\\
.&.&.&.&.&1&.&.\\
\end{array} \right),
\end{equation}
\begin{equation}
e_{R1} \rightarrow  \left( \begin{array} {cccccccc}
.&-1&.&.&.&.&.&.\\
1&.&.&.&.&.&.&.\\
.&.&.&.&.&.&1&.\\
.&.&.&.&1&.&.&.\\
.&.&.&-1&.&.&.&.\\
.&.&.&.&.&.&.&1\\
.&.&-1&.&.&.&.&.\\
.&.&.&.&.&-1&.&.\\
\end{array} \right)
\end{equation}
(only nonzero entries are indicated).
Note that $e_{2}e_{6}=e_{3}e_{4}=e_{5}e_{7}=e_{1}$,
but because of the nonassociativity of {\bf O},
for example,
\begin{equation}
e_{1}x = (e_{2}e_{6})x \ne e_{2}(e_{6}x) \equiv e_{L26}(x)
\end{equation}
in general.
\newline
\vspace{12pt}

In the octonion algebra, any product from the right can
be reproduced as the sum
of products from the left, and visa versa.
\newline
\vspace{12pt}

The left and right adjoint algebras ${\bf O}_{L}$
and ${\bf O}_{R}$ are the same algebra, and this algebra
is larger than {\bf O} itself.
In fact, it is isomorphic to {\bf R}(8).

\vspace{12pt}

It is not difficult to prove that
\begin{equation}
e_{La...bc...d}= -e_{La...cb...d}
\end{equation}
if $b \neq c$, all indices from 1 to 7.

\vspace{12pt}

So,for example, $e_{1}(e_{2}(e_{3}x))  = -e_{1}(e_{3}(e_{2}x))
= e_{3}(e_{1}(e_{2}x))$.

\vspace{12pt}

In addition,
\begin{equation}
e_{Lab...pp...c} = -e_{Lab...c}
\end{equation}
(cancellation of like indices).

\vspace{12pt}

Together with
\begin{equation}
e_{L7654321} = {\bf 1},
\end{equation}
Equations (16) and (17) imply that a complete basis for
the left=right adjoint algebra
of {\bf O} consists of elements of the form
\begin{equation}
{\bf 1}, e_{La}, e_{Lab}, e_{Labc},
\end{equation}
{\bf 1} the identity.
\vspace{12pt}

This yields 1+7+21+35=64 as the
dimension of the adjoint
algebra of {\bf O}, also the dimension of {\bf R}(8).
\begin{equation}
{\bf O}_{L}={\bf O}_{R}={\bf R}(8).
\end{equation}
The 8-dimensional {\bf O} itself is the object space
of the adjoint algebra.

\newpage

\subsection{Aut(O) and Der(O).}
The 14-dimensional exceptional Lie group $G_{2}$ is
the automorphism group
$Aut(\bf{O})$ of the octonions.
\newline
\vspace{12pt}

When viewed from the point of view of a linear
Lie algebra with bracket product
rather than the point of view of a nonlinear
global Lie group with group product,
the structure that corresponds to the
automorphism group $Aut(\bf{O})$ is the
algebra of derivations $Der(\bf{O})$ of the octonions.
\vspace{12pt}

A derivation is a linear map $D:\bf{O} \rightarrow \bf{O}$
such that for $x,y \in \bf{O}$:
\begin{equation}
D(xy) = (Dx)y + x(Dy)
\end{equation}
Then, from the alternative property of the octonions:
\begin{equation}
D(ab)x = [a,b,x] + (1/3)[[a,b],x]
\end{equation}

\vspace{12pt}

Let $C_{d} = L_{d} - R_{d}$, where $L_{d}$ and $R_{d}$
denote left and right
multiplication by an imaginary octonion $d$.
The imaginary octonions $Im(\bf{O})$
are those octonions in the space orthogonal
to the octonion real axis.  Then
$$D(ab) = (1/6)([C_{a},C_{b}] + C_{[a,b]})$$
\vspace{12pt}

The algebra of derivations $Der(\bf{O})$ of
the octonions is the Lie algebra $G_{2}$.
\newline
\vspace{12pt}

A basis for the Lie algebra of $G_{2}$ is represented
in ${\bf O}_{L}$ by:
\begin{equation}
\{e_{Lab}-e_{Lcd} : e_{a}e_{b} = e_{c}e_{d}\} \rightarrow G_{2}.
\end{equation}
In ${\bf O}_{R}$ the basis is much the same:
$$
\{e_{Rab}-e_{Rcd} : e_{a}e_{b} = e_{c}e_{d}\} \rightarrow G_{2}.
$$
The stability group of any fixed octonion direction is the
8-dimensional $SU(3) \subset G_{2}$.
\vspace{12pt}

A basis for the Lie algebra of the stability group of $e_{7}$ is:
\begin{equation}
\{e_{Lab}-e_{Lcd} \in G_{2} : a,b,c,d \ne 7\} \rightarrow su(3).
\end{equation}
Thus $SU(3)$ is the intersection of $G_{2}$ with $Spin(6)$, and
\newline
$G_{2} = SU(3) \oplus S^{6}$.
\newline
\vspace{12pt}

The algebra of derivations does not give the Lie algebra
all the antisymmetric maps of a real division algebra
$\bf{R, C, H, O}$ unless the algebra is commutative and
associative, i.e., $\bf{R, C}$.  In the case of $\bf{C}$,
the Lie algebra of all antisymmetric maps is
$$Der({\bf{C}}) = Spin(2) = U(1)$$.
\vspace{12pt}

For $\bf{H}$, which is associative but not commutative,
$$Der({\bf{H}}) \subset L_{Im({\bf{H}})} \oplus R_{Im({\bf{H}})}$$
where $L_{Im({\bf{H}})}$ and $R_{Im({\bf{H}})}$ denote left and
right multiplication on the quaternions by the imaginary quaternions.
\newline
$L_{Im({\bf{H}})}$ is isomorphic to and commutes
with $R_{Im({\bf{H}})}$.
\newline
Since $L_{Im({\bf{H}})} = R_{Im({\bf{H}})} =
Spin(3) = SU(2) = Sp(1) = S^{3}$,
the Lie algebra $Spin(4)$ of all antisymmetric maps of
$\bf{H}$ is given by
$$Spin(4) = Spin(3) \oplus Spin(3)$$
\vspace{12pt}

For $\bf{O}$, which is neither associative nor commutative,
the vector space of all antisymmetric maps of $\bf{O}$
is given by
$$Spin(0,8) = Der({\bf{O}}) \oplus L_{Im({\bf{O}})} \oplus
R_{Im({\bf{O}})} =
G_{2} \oplus S^{7} \oplus S^{7}$$
where $S^{7}$ represents the imaginary octonions,
notation suggested by the fact that the unit octonions
are the 7-sphere $S^{7}$ which, since it is parallelizable,
is locally representative of the imaginary octonions.
\vspace{12pt}

The 7-sphere $S^{7}$ has a left-handed basis $\{e_{La}\}$ and
\newline
a right-handed basis $\{e_{Ra}\}$.
\newline
\vspace{12pt}

Unlike the 3-sphere, which is the Lie group $Spin(3) = SU(2)
= Sp(1)$,
\newline
$S^{7}$ does not close under the commutator bracket product because
\newline
$[e_{La},e_{Lb}]/2 = e_{Lab}$ and $[e_{Ra},e_{Rb}]/2 = e_{Rab}$
for $a \neq b$.
\newline
\vspace{12pt}

To make a Lie algebra out of $S^{7}$, it must be extended
to $Spin(0,8)$ by adding
the 21 basis elements $\{e_{Lab}\}$ or $\{e_{Rab}\}$.
\newline
\vspace{12pt}

There are three ways to extend $S^{7}$ to the Lie algebra
$Spin(0,8)$.
\newline
\vspace{12pt}

They result in the left half-spinor, right half-spinor,
and vector representations
of $Spin(0,8)$:
\begin{equation}
\begin{array}{c}
\{e_{La},e_{Lbc}\} \rightarrow Spin(0,8), \mbox{left half-spinor}, \\
\{e_{Ra},-e_{Rbc}\} \rightarrow Spin(0,8), \mbox{right half-spinor}, \\
\{e_{La}+e_{Ra}, e_{Lbc} - e_{Ra}:e_{a}=e_{b}e_{c}\} \rightarrow
Spin(0,8), \mbox{vector}. \\
\end{array}
\end{equation}
Therefore, the two half-spinor representations and
the vector representation of $Spin(0,8)$ all have 8-dimensional {\bf O}
as representation space.
\vspace{12pt}

The three representations are isomorphic
by triality.
\vspace{12pt}

Consider the 28 basis elements $\{e_{La}+e_{Ra}, e_{Lbc} - e_{Ra}:
e_{a}=e_{b}e_{c}\}$ of
the vector representation of $Spin(0,8)$:
\newline
The 21-element subset $\{e_{Lbc} - e_{Ra}:e_{a}=e_{b}e_{c}\}$
is a basis
\newline
for the Lie algebra $Spin(7) = G_{2} \oplus S^{7}$.
\newline
\vspace{12pt}

Therefore, the Lie algebra $Spin(0,8) = S^{7} \oplus G_{2}
\oplus S^{7}$.

\newpage

\section{$3 \times 3$ Octonion Matrices and $E_{6}$.}
The $D_{4}-D_{5}-E_{6}$ model of physics uses 3 copies
of the octonions:
\newline
$\bf O_{v}$ to represent an 8-dimensional spacetime
(prior to dimensional reduction to 4 dimensions);
\newline
$\bf O_{+}$ to represent the 8 first-generation
fermion +half-spinor particles ; and
\newline
$\bf O_{-}$ to represent the 8 first-generation
fermion -half-spinor antiparticles.
\newline
\vspace{12pt}

Consider the 72-dimensional space of $3 \times 3$ matrices of
octonions:
\begin{equation}
\left(
\begin{array}{ccc}
{\bf O}_{1} & {\bf O}_{+} &  {\bf O}_{v} \\
& & \\
{\bf X} & {\bf O}_{2} &  {\bf O}_{-} \\
& & \\
{\bf Z} & {\bf Y} & {\bf O}_{3}
\end{array}
\right)
\end{equation}
where
\newline
${\bf O}_{v}, {\bf O}_{+}, {\bf O}_{-},{\bf O}_{1}, {\bf O}_{2},
{\bf O}_{3}, {\bf X}, {\bf Y}, {\bf Z}$ are octonion,
\newline
${\bf O}_{1} = a + S^{7}_{1}, {\bf O}_{2} = b + S^{7}_{2},
{\bf O}_{3} =  c + S^{7}_{3}$
\newline
$a, b, c$ are real, and
\newline
$S^{7}_{1}, S^{7}_{2}, S^{7}_{3}$ are imaginary octonion.
\newline
\vspace{12pt}

Consider the ordinary matrix product $AB$
of two $3 \times 3$ octonion matrices $A$ and $B$.
\newline
\vspace{12pt}

Now, to construct the Lie algebra $E_{6}$
from $3 \times 3$ octonion matrices, it is useful to
split the product $AB$ into antisymmetric and symmetric parts.
\begin{equation}
AB = (1/2)(AB - BA) + (1/2)(AB + BA)
\end{equation}
\vspace{12pt}

This will enable us to construct an $F_{4}$ Lie algebra
from antiHermitian matrices that will arise from
considering the antisymmetric product, and
\newline
a $J_{3}^{\bf O}o$ Jordan algebra from hermitian matrices
that will arise from considering the symmetric product.
\newline
\vspace{12pt}

Then the Lie algebra $F_{4}$ and the Jordan algebra
$J_{3}^{\bf O}o$ will be combined to form the
Lie algebra $E_{6}$ whose structure forms the basis
of the $D_{4}-D_{5}-E_{6}$ model.

\vspace{12pt}

This construction is only possible in this case
because of many exceptional structures and symmetries.
The $D_{4}-D_{5}-E_{6}$ model therefore inherits
remarkable symmetry structures

\newpage

\subsection{Antihermitian Matrices and Lie Algebras.}
Consider the antisymmetric product $(1/2)(AB - BA)$:
\newline
The 45-dimensional space of antihermitian $3 \times 3$
octonion matrices does not close under the antisymmetric
product to form a Lie algebra
(here, $\dagger$ denotes octonion conjugation):
\begin{equation}
\left(
\begin{array}{ccc}
S^{7}_{1} & {\bf O}_{+} &  {\bf O}_{v} \\
& & \\
-{\bf O}_{+}^\dagger & S^{7}_{2} &  {\bf O}_{-} \\
& & \\
-{\bf O}_{v}^\dagger & -{\bf O}_{-}^\dagger & S^{7}_{3}
\end{array}
\right)
\end{equation}
A product that closes is
\begin{equation}
(1/2)(AB - BA - Tr(AB - BA))
\end{equation}
\newline
The form of the product indicates that to
get closure, you have to use only the 45-7 = 38-dimensional
space of traceless antihermitian $3 \times 3$ octonion matrices:
\begin{equation}
\left(
\begin{array}{ccc}
S^{7}_{1} & {\bf O}_{+} &  {\bf O}_{v} \\
& & \\
-{\bf O}_{+}^\dagger & S^{7}_{2} &  {\bf O}_{-} \\
& & \\
-{\bf O}_{v}^\dagger & -{\bf O}_{-}^\dagger & -S^{7}_{1}-S^{7}_{2}
\end{array}
\right)
\end{equation}
However, the Jacobi identity is not satisfied, so you still
do not have a Lie algebra.
\newline
\vspace{12pt}

As was mentioned in Section 1.3, to get the Lie algebra of
all antisymmetric maps of the nonassociative, noncommutative
real division algebra ${\bf{O}}$ you must include the 14-dimensional
Lie algebra of derivations $Der({\bf{O}}) = G_{2}$.
\newline
\vspace{12pt}

Adding the derivations to the product that closes gives a
product that not only closes but also satisfies the Jacobi identity:
\begin{equation}
(1/2)(AB - BA - Tr(AB -BA) \oplus D(A,B))
\end{equation}
where the derivation $D(A,B) = \sum_{ij} D(a_{ij},b_{ij})$
\newline
and the derivation $D(x,y)$ acts on the octonion $z$
\newline
by using the alternator $[x,y,z] = D(x,y)z$.
\newline
\vspace{12pt}

The resulting space is the 38+14 = 52-dimensional Lie algebra $F_{4}$.
\begin{equation}
\left(
\begin{array}{ccc}
S^{7}_{1} & {\bf O}_{+} &  {\bf O}_{v} \\
& & \\
-{\bf O}_{+}^\dagger & S^{7}_{2} &  {\bf O}_{-} \\
& & \\
-{\bf O}_{v}^\dagger & -{\bf O}_{-}^\dagger & -S^{7}_{1}-S^{7}_{2}
\end{array}
\right)
\oplus G_{2}
\end{equation}
The physical interpretation of this representation of $F_{4}$ in
the $D_{4}-D_{5}-E_{6}$ model is:
\vspace{12pt}

${\bf O}_{v}$ is 8-dimensional spacetime before dimensional
reduction to 4 dimensions;
\vspace{12pt}

${\bf O}_{+}$ is the 8-dimensional space representing the first
generation fermion particles (the neutrino, the electron,
the red, blue and green up quarks, and the red, blue and green
down quarks);
\vspace{12pt}

${\bf O}_{-}$ is the 8-dimensional space representing the
first generation fermion antiparticles
before dimensional reduction creates 3 generations; and
\vspace{12pt}

$S^{7}_{1} \oplus S^{7}_{2} \oplus G_{2}$ is the $Spin(0,8)$
gauge group before dimensional
reduction breaks it down into gravity plus the Standard Model.

\vspace{12pt}

Now, consider the symmetric Jordan product and
the hermitian matrices that form an algebra under it.

\vspace{12pt}

In the next subsection, they will be studied so that
their structure can be added to the $F_{4}$ structure of traceless
antihermitian matrices plus the derivations $G_{2}$.
\vspace{12pt}

\newpage

\subsection{Hermitian Matrices and Jordan Algebras.}
Consider the symmetric product $(1/2)(AB + BA)$:
\newline
\vspace{12pt}

The 27-dimensional space of hermitian $3 \times 3$ octonion
matrices closes under the symmetric product to form the Jordan
algebra $J_{3}^{O}$:
\begin{equation}
\left(
\begin{array}{ccc}
a & {\bf O}_{+} &  {\bf O}_{v} \\
& & \\
{\bf O}_{+}^\dagger & b &  {\bf O}_{-} \\
& & \\
{\bf O}_{v}^\dagger & {\bf O}_{-}^\dagger & c
\end{array}
\right)
\end{equation}
Even though the full 27-dimensional space of hermitian matrices
forms a Jordan algebra $J_{3}^{O}$, only the 26-dimensional
traceless subalgebra $J_{3}^{O}o$ is acted on by $F_{4}$ as
its representation space:
\begin{equation}
\left(
\begin{array}{ccc}
a & {\bf O}_{+} &  {\bf O}_{v} \\
& & \\
{\bf O}_{+}^\dagger & b &  {\bf O}_{-} \\
& & \\
{\bf O}_{v}^\dagger & {\bf O}_{-}^\dagger & -a-b
\end{array}
\right)
\end{equation}
The 52-dimensional $F_{4}$ and the 26-dimensional $J_{3}^{O}o$
combine to form the 78-dimensional Lie algebra $E_{6}$,

\vspace{12pt}

\begin{equation}
\left(
\begin{array}{ccc}
S^{7}_{1} & {\bf O}_{+} &  {\bf O}_{v} \\
& & \\
-{\bf O}_{+}^\dagger & S^{7}_{2} &  {\bf O}_{-} \\
& & \\
-{\bf O}_{v}^\dagger & -{\bf O}_{-}^\dagger & -S^{7}_{1}-S^{7}_{2}
\end{array}
\right)
\oplus G_{2} \oplus
\left(
\begin{array}{ccc}
a & {\bf O}_{+} &  {\bf O}_{v} \\
& & \\
{\bf O}_{+}^\dagger & b &  {\bf O}_{-} \\
& & \\
{\bf O}_{v}^\dagger & {\bf O}_{-}^\dagger & -a-b
\end{array}
\right)
\end{equation}

the $E_{6}$ of the $D_{4}-D_{5}-E_{6}$ model.

\vspace{12pt}

$E_{6}$ preserves the cubic determinant pseudoscalar
3-form for $3 \times 3$ octonionic matrices
(see \cite{FUL,CHE}).
\newline
\vspace{12pt}

Sudbery \cite{SUD} has identified $E_{6}$ with $SL(3,{\bf O})$.
\newline
\vspace{12pt}

Flynn \cite{FLY}, in the context of her physics models,
\newline
has used such an identification to note
\newline
the similarity of $E_{6}$ to $SL(3,{\bf C})$,
\newline
which is also made up of an antisymmetric part, $SU(3)$,
\newline
plus a symmetric part, an 8-dimensional Jordan algebra,
\newline
and which preserves a cubic determinant.

\vspace{12pt}

\newpage

\section{Shilov Boundaries of Complex Domains.}

In the $D_{4}-D_{5}-E_{6}$ model, physical spacetime and
the physical spinor fermion representation manifold are
Shilov boundaries of bounded complex domains.
\vspace{12pt}

The best general reference to Shilov boundaries is
Helgason's 1994 book \cite{HEL2}.
\vspace{12pt}

The best set of calculations of volumes, etc., of
Shilov boundaries is Hua's book \cite{HUA}.

\vspace{12pt}

That means, for instance, that physical 8-dimensional spacetime
is the 8-real dimensional Shilov boundary of of a
16-real dimensional (8-complex-dimensional) bounded complex
domain.

\vspace{12pt}

To physicists, the most familiar example of bounded complex
domains and their Shilov boundaries (other than the unit disk
and its Shilov boundary, the circle) probably comes from
the twistors of Roger Penrose \cite{PEN}.
\newline
Another example, possibly less familiar, is
the chronometry theory of I. E. Segal \cite{SEG} at M.I.T.
\newline
Still another example, also probably not very familiar,
is the use of the geometry of bounded complex domains
by Armand Wyler \cite{WYL} in his effort to calculate
the value of the fine structure constant.

\vspace{12pt}

To mathematicians, such structures are well known.
\newline
A standard general reference is the book of Hua \cite{HUA},
in which Shilov boundaries are called characteristic manifolds.
\newline
Actually, I would prefer the term characteristic boundary,
because it would describe it as being part of a boundary
that characterizes important structures on the manifold.
\newline
However, I will use the term Shilov boundary because that
seems to be the dominant term in English-language literature.

\vspace{12pt}

The simplest example (a mathematical object that Prof. Feller
\cite{FEL} said was the best all-purpose example in mathematics
for understanding new concepts) is the unit disk along with its
harmonic functions.  The unit disk is the bounded complex
domain, the unit circle is its Shilov boundary, and the
harmonic functions are determined throughout the unit
disk by their values on the Shilov boundary.

\vspace{12pt}

A more complicated example of such structures,
\newline
taken from the works of those mentioned above,
\newline
starts with an 8-real-dimensional 4-complex-dimensional
\newline
space denoted ${\cal M}^{\bf C}$ with signature $(2,6)$.

\vspace{12pt}

{\bf What are Bounded Complex Homogeneous Domains?}

\vspace{12pt}

To see the second example, start with the complexified
Minkowski spacetime ${\cal M}_{2,6}^{\bf C}$
of Penrose twistor theory.

\vspace{12pt}

${\cal M}^{\bf C}$ is a Hermitian symmetric space
that is the coset space $$Spin(2,4) / Spin(1,3) \times U(1)$$.
In the mathematical classification notation, it is called a
\newline
Hermitian symmetric space of type $BDI_{2,4}$.

\vspace{12pt}

The Hermitian symmetric coset space is unbounded, but
for each Hermitian symmetric space there exists a
natural corresponding bounded complex domain.

\vspace{12pt}

In this case, $BCI_{2,4}$, the bounded complex domain is
called $Type \; IV_{4}$ and consists of the elements of
${\bf C}_{4}$ defined by
\begin{equation}
\{ z_{1} ,z_{2}, z_{3} ,z_{4} \mid
|z_{1}|^{2} + |z_{2}|^{2} + |z_{3}|^{2} + |z_{4}|^{2} <
(1 + |z_{1}^{2} + z_{2}^{2} + z_{3}^{2} + z_{4}^{2}|) / 2 < 1 \}
\end{equation}

\vspace{12pt}

{\bf What are Shilov boundaries?}

\vspace{12pt}

The Shilov boundary (called the characteristic manifold
by Hua in \cite{HUA}) is a subset of the topological boundary
of the bounded complex domain.

\vspace{12pt}

Following Hua \cite{HUA}, consider the analytic functions on the
bounded complex domain.  The Shilov boundary is the part of
the topological boundary on which every analytic function attains
its maximum modulus, and such that for every point on the Shilov
boundary, there exists an analytic function on the bounded complex
domain that attains its maximum modulus at that point.

\vspace{12pt}

The Shilov boundary is closed.  Any function which is analytic in
the neighborhood of every point of the Shilov boundary is uniquely
determined by its values on the Shilov boundary.

\vspace{12pt}

In the case of the bounded domain of $Type \; IV_{4}$ Hua \cite{HUA}
shows that the Shilov boundary is
\begin{equation}
\{ z = e^{i \theta} x \mid 0 \leq \theta \leq \pi,
x \overline{x} = 1 \} = {\bf R}P^{1} \times S^{3}
\end{equation}

In the Penrose twistor formalism, it is the
4-dimensional Minkowski spacetime ${\cal M}_{1,3}$
with signature $(1,3)$

\vspace{12pt}

There exists a kernel function, the Poisson kernel
$P(z, \xi)$ function of a point z in the
bounded complex domain and a point $\xi$ in its Shilov
boundary,  such that, for any analytic $f(z)$,
\begin{equation}
f(z) = \int_{Shilov bdy} P(z, \xi) f(\xi)
\end{equation}
Since all the analytic functions in the bounded complex
domain are determinmed by their values on the Shilov
boundary,the Shilov boundary should be the proper domain
of definition of physically relevant functions.
\newline
That is why the $D_{4}-D_{5}-E_{6}$ model takes the
Shilov boundary to be the relevant manifold for spacetime and
for representing fermion particles and antiparticles.

\vspace{12pt}

There exists another kernel function,
\newline
the Bergman kernel function $K(z, \overline{w})$
\newline
of two points $z, w$ in the bounded complex domain
\newline
such that, for any analytic $f(z)$,
\begin{equation}
f(z) = \int_{domain} K(z, \overline{w}) f(w)
\end{equation}

\vspace{12pt}

Setting $z = w$ in the Bergman kernel gives a Riemannian metric
for the bounded complex domain,
\newline
which in turn defines invariant differential operators including
the Laplacian,
\newline
which in turn gives harmonic functions.

\vspace{12pt}

The Bergman kernel is equal to the ratio of the volume density
to the Euclidean volume of the bounded complex domain.

\vspace{12pt}

Hua \cite{HUA} not only gives the above description, he also
actually calculates the volumes of the bounded complex domains
and their Shilov boundaries.

\vspace{12pt}

Suppose that, as in the $D_{4}-D_{5}-E_{6}$ model, different
bounded complex domains represent different physical forces whose
Green's functions are determined by their invariant differential
operators.
\newline
Then, since the domain volumes represent the measures of the
Bergman kernels of bounded complex domains, and
\newline
since the physical part of the domain is its Shilov boundary,
the ratios of (suitably normalized) volumes of Shilov boundaries
should (and do, in the $D_{4}-D_{5}-E_{6}$ model) represent
the ratios of force strengths of the corresponding forces.

\vspace{12pt}

\newpage

\section{$E_{6} / (D_{5} \times U(1))$ and $D_{5} / (D_{4} \times U(1))$.}
\subsection{2x2 Octonion Matrices and D5.}
The $3 \times 3$ octonion traceless antihermitian checkerboard
matrices form a subalgebra of the $3 \times 3$ octonion
traceless antihermitian matrices.
\begin{equation}
\left(
\begin{array}{ccc}
S^{7}_{1} & 0 &  {\bf O}_{v} \\
& & \\
0 & S^{7}_{2} &  0 \\
& & \\
-{\bf O}_{v}^\dagger & 0 & -S^{7}_{1}-S^{7}_{2}
\end{array}
\right)
\end{equation}
It is isomorphic to the algebra of $2 \times 2$ octonion
antihermitian matrices:
\begin{equation}
\left(
\begin{array}{cc}
S^{7}_{1} &  {\bf O}_{v} \\
& \\
-{\bf O}_{v}^\dagger & S^{7}_{2}
\end{array}
\right)
\end{equation}
When $Der({\bf{O}}) = G_{2}$ is added the result is a
subalgebra of the Lie algebra $F_{4}$:
\begin{equation}
\left(
\begin{array}{cc}
S^{7}_{} &  {\bf O}_{v} \\
& \\
-{\bf O}_{v}^\dagger & S^{7}_{2}
\end{array}
\right)
\oplus G_{2}
\end{equation}
This Lie algebra is 8+7+7+14 = 36-dimensional $Spin(9)$,
also denoted $B_{4}$.
\newline
$Spin(9)$ is to ${\bf{O}}$ as $SU(2)$ is to ${\bf{C}}$.
\vspace{12pt}

The checkerboard $3 \times 3$ octonion traceless hermitian
matrices also form a subalgebra of the  $3 \times 3$ octonion
traceless hermitian matrices:
\begin{equation}
\left(
\begin{array}{ccc}
a & 0 &  {\bf O}_{v} \\
& & \\
0 & b &  0 \\
& & \\
{\bf O}_{v}^\dagger & 0 & -a-b
\end{array}
\right)
\end{equation}
It is isomorphic to the 10-dimensional Jordan algebra
$J_{2}^{\bf{O}}$ of $2 \times 2$ octonion hermitian matrices:
\begin{equation}
\left(
\begin{array}{cc}
a &  {\bf O}_{v} \\
& \\
{\bf O}_{v}^\dagger & b
\end{array}
\right)
\end{equation}
$J_{2}^{\bf{O}}$ has a 9-dimensional traceless Jordan
subalgebra $J_{2}^{\bf{O}}o$, and
$$J_{2}^{\bf{O}} = J_{2}^{\bf{O}}o \oplus U(1)$$
The 1-dimensional Jordan algebra $J_{1}^{\bf{C}}o$ corresponds
to the Lie algebra $U(1)$.
\vspace{12pt}

The 36-dimensional $B_{4}$ and the 9-dimensional
$J_{2}^{\bf{O}}o$ combine to form the 45-dimensional
Lie algebra $D_{5}$, also denoted $Spin(10)$,
the $D_{5}$ of the $D_{4}-D_{5}-E_{6}$ model.
\vspace{12pt}

Physically, $D_{5}$ contains the gauge group and
spacetime parts of $E_{6}$,
so the half-spinor fermion particle and
antiparticle parts of $E_{6}$ should
live in a coset space that is a quotient
of $E_{6}$ by $D_{5}$.
\vspace{12pt}

Due to complex structure, the quotient must be taken
by $D_{5} \times U(1)$ rather than $D_{5}$ alone.
The $U(1)$ comes from action on the Jordan
algebra  $J_{1}^{\bf{C}}o$.
\vspace{12pt}

The resulting symmetric space that is the representation
space of the first-generation particles and antiparticles
in the $D_{4}-D_{5}-E_{6}$ model is the 78-45-1 =
32-real-dimensional hermitian symmetric space
\begin{equation}
E_{6} / (D_{5} \times U(1))
\end{equation}
The 16 particles and antiparticles live on the
16-real-dimensional Shilov boundary
of the bounded complex domain that corresponds to
the symmetric space.
The Shilov boundary is
\begin{equation}
(S^{7} \times {\bf{R}}P^{1}) \oplus  (S^{7} \times {\bf{R}}P^{1})
\end{equation}
There is one copy of $S^{7} \times {\bf{R}}P^{1}$ for
the 8 first-generation fermion
half-spinor particles, and one for the 8 antiparticles.
\vspace{12pt}

\newpage

\subsection{2x2 Diagonal Octonion Matrices and D4.}
The $2 \times 2$ octonion antihermitian checkerboard matrices
form the diagonal matrix subalgebra of the $2 \times 2$ octonion
antihermitian matrices.
\begin{equation}
\left(
\begin{array}{cc}
S^{7}_{1} & 0\\
& \\
0 & S^{7}_{2}
\end{array}
\right)
\end{equation}
When $Der({\bf{O}}) = G_{2}$ is added the result is a
subalgebra of the Lie algebras $B_{4}$ and $F_{4}$:
\begin{equation}
\left(
\begin{array}{cc}
S^{7}_{1} & 0\\
& \\
0 & S^{7}_{2}
\end{array}
\right)
\oplus G_{2}
\end{equation}
This Lie algebra is 7+7+14 = 28-dimensional $Spin(0,8)$,
also denoted $D_{4}$.
\newline
\vspace{12pt}

The checkerboard $2 \times 2$ octonion traceless hermitian
matrices also form the diagonal subalgebra of the  $2 \times 2$
octonion traceless hermitian matrices:
\begin{equation}
\left(
\begin{array}{cc}
a & 0\\
& \\
0 & 0
\end{array}
\right)
\end{equation}
It is isomorphic to the 1-dimensional Jordan algebra
$J_{1}^{\bf{C}}o$.
\newline
\vspace{12pt}

The 1-dimensional Jordan algebra $J_{1}^{\bf{C}}o$ corresponds
to the Lie algebra $U(1)$.
\vspace{12pt}

 The 28-dimensional $D_{4}$ Lie algebra, also denoted $Spin(0,8)$,
is (before dimensional reduction) the smallest Lie algebra in
the $D_{4}-D_{5}-E_{6}$ model.
\vspace{12pt}

Physically, $D_{4}$ contains only the gauge group parts of
$D_{5}$ and $E_{6}$, so the 8-dimensional spacetime of $D_{5}$ and
$E_{6}$ should live in a coset space that is a quotient of $D_{5}$
by $D_{4}$.
\vspace{12pt}

Due to complex structure, the quotient must be taken by
$D_{4} \times U(1)$ rather than $D_{4}$ alone.
The $U(1)$ comes from action on the Jordan algebra  $J_{1}^{\bf{C}}o$.
\vspace{12pt}

The resulting symmetric space that is the representation space of
the 8-dimensional spacetime in the $D_{4}-D_{5}-E_{6}$ model is
the 45-28-1 = 16-real-dimensional hermitian symmetric space
\begin{equation}
D_{5} / (D_{4} \times U(1))
\end{equation}
The physical (before dimensional reduction) spacetime lives on
the 8-real-dimensional Shilov boundary of the bounded complex domain
that corresponds to the symmetric space.  The Shilov boundary is
\begin{equation}
S^{7} \times {\bf{R}}P^{1}
\end{equation}
\vspace{12pt}

\newpage

\section{Global $E_{6}$ Lagrangian.}

Represent the $E_{6}$ Lie algebra by

\begin{equation}
{\bf R} \otimes
\left(
\left(
\begin{array}{ccc}
S^{7}_{1} & {\bf O}_{+} &  {\bf O}_{v} \\
& & \\
-{\bf O}_{+}^\dagger & S^{7}_{2} &  {\bf O}_{-} \\
& & \\
-{\bf O}_{v}^\dagger & -{\bf O}_{-}^\dagger & -S^{7}_{1}-S^{7}_{2}
\end{array}
\right)
\oplus G_{2} \oplus
\left(
\begin{array}{ccc}
a & {\bf O}_{+} &  {\bf O}_{v} \\
& & \\
{\bf O}_{+}^\dagger & b &  {\bf O}_{-} \\
& & \\
{\bf O}_{v}^\dagger & {\bf O}_{-}^\dagger & -a-b
\end{array}
\right)
\right)
\end{equation}

where the {\bf R} represents the real scalar field
of this representation of $E_{6}$.

Full $E_{6}$ symmetry of the $D_{4}-D_{5}-E_{6}$ model is
a global symmetry,
and is useful primarily to:
\newline
define the spacetime, fermions, and bosons;
\newline
define the tree-level relative force strengths and particle masses;
\newline
give generalized supersymmetric relationships among fermions and
bosons that may be useful in loop cancellations to produce
ultraviolet and infrared finite results;
\newline
give $CPT$ symmetry, which involves both particle $C$ and
spacetime $P$ and $T$ symmetries.
\newline
Some more discussion of $CPT$, $CP$, and $T$ can be found at
\newline
\href{http://www.gatech.edu/tsmith/CPT.html}{WWW
URL http://www.gatech.edu/tsmith/CPT.html} \cite{SMI6}.

\vspace{12pt}

Dynamics of the $D_{4}-D_{5}-E_{6}$ model are given by a
Lagrangian action that is the integral over spacetime of
a Lagrangian density made up of
a gauge boson curvature term, a spinor fermion term
(including through a Dirac operator interaction with gauge bosons),
and a scalar term.
\newline
\vspace{12pt}

The 8-dimensional Lagrangian is:

\begin{equation}
\int_{V_{8}} F_{8} \wedge \star F_{8} + \partial_{8}^{2}
\overline{\Phi_{8}} \wedge \star \partial_{8}^{2} \Phi_{8} +
\overline{S_{8\pm}} \not \! \partial_{8} S_{8\pm}
\end{equation}

where $\star$ is the Hodge dual;
\newline
\vspace{12pt}

$\partial_{8}$ is the 8-dimensional covariant derivative,
\newline
$\not \!  \partial_{8}$ is the 8-dimensional Dirac operator,
and
\newline
$F_{8}$ is the 28-dimensional $Spin(8)$ curvature,
\newline
which come from the $Spin(0,8)$ gauge group subgroup
of $E_{6}$, represented here by

\begin{equation}
\left(
\begin{array}{cc}
S^{7}_{1} & 0  \\
&  \\
0 & S^{7}_{2}
\end{array}
\right)
\oplus G_{2}
\end{equation}

\vspace{12pt}

$\Phi_{8}$ is the 8-dimensional scalar field,
which comes from the ${\bf R}$ scalar part of the
representation of $E_{6}$;
\newline
\vspace{12pt}

$V_{8}$ is 8-dimensional spacetime, and
which comes from the ${\bf O}_{v}$ part of the
representation of $E_{6}$;
\newline
\vspace{12pt}

$S_{8\pm}$ are the $+$ and $-$ 8-dimensional
half-spinor fermion spaces,
which come from the ${\bf O}_{+}$ and ${\bf O}_{-}$
parts of the representation of $E_{6}$;
\newline
\vspace{12pt}

The 8-dimensional Lagrangian is a classical Lagrangian.
\newline
\vspace{12pt}

To get a quantum action, the $D_{4}-D_{5}-E{6}$ model uses
a path integral sum over histories.  \newline
At its most fundamental level, the $D_{4}-D_{5}-E{6}$ model is a
lattice model with 8-dimensional $E_{8}$ lattice spacetime.
\newline
\vspace{12pt}

The path integral sum over histories is based on a generalized
Feynman checkerboard scheme over the $E_{8}$ lattice spacetime.
\newline
\vspace{12pt}

The original Feynman checkerboard is in 2-dimensional spacetime,
in which the speed of light is naturally $c$ = $1$.
\newline
\vspace{12pt}

In the 4-dimensional spacetime of the $D_{4}$ lattice,
the speed of light is naturally $c$ = $\sqrt{3}$.
\newline
\vspace{12pt}

In 8-dimensional spacetime, the speed of light is
naturally $c$ = $\sqrt{7}$, but in the 8-dimensional $E_{8}$
lattice the nearest neighbor vertices have only 4 (not 8)
non-zero coordinates and therefore have a natural speed of
light of $c$ = $\sqrt{3}$ that is appropriate for 4-dimensional
light-cones rather than for 8-dimensional lightcones with
$c$ = $\sqrt{7}$.
\vspace{12pt}

This means that:
\newline
\vspace{12pt}

the generalized Feynman checkerboard scheme will not produce
lightcone paths in 8-dimensional spacetime;
\newline
\vspace{12pt}

the dimension of spacetime must be reduced to 4 dimensions,
as discussed in Section 7.; and
\newline
the 8-dimensional Lagrangian is classical, and does not contain
gauge fixing and ghost terms.
\newline
\vspace{12pt}

Gauge-fixing term and ghost terms only appear
after dimensional reduction of spacetime
permits construction of generalized Feynman checkerboard
quantum path integral sums over histories.
\newline
\vspace{12pt}

The gauge fixing terms are needed to avoid overcounting
gauge-equivalent paths, and ghosts then appear.
\newline
\vspace{12pt}

The $D_{4} = Spin(0,8)$  gauge symmetry will be reduced to gravity
plus the Standard Model by dimensional reduction.
The gauge symmetries of gravity and the Standard Model will be
broken by gauge-fixing and ghost terms, and
the resulting Lagrangian will have a BRS symmetry.
\vspace{12pt}

The dynamical symmetry that is physically relevant at
the initial level is the $D_{4} = Spin(0,8)$ gauge symmetry
of the Lagrangian.
\vspace{12pt}

It is clear that $Spin(0,8)$ acts through its 8-dimensional
vector, +half-spinor, and -half-spinor representations on
the ${\bf O}_{v}$, ${\bf O}_{+}$, and
${\bf O}_{-}$ parts of the antihermitian matrix
\begin{equation}
\left(
\begin{array}{ccc}
S^{7}_{1} & {\bf O}_{+} &  {\bf O}_{v} \\
& & \\
-{\bf O}_{+}^\dagger & S^{7}_{2} &  {\bf O}_{-} \\
& & \\
-{\bf O}_{v}^\dagger & -{\bf O}_{-}^\dagger & -S^{7}_{1}-S^{7}_{2}
\end{array}
\right)
\end{equation}
\vspace{12pt}

It is also clear that $Spin(0,8)$ acts through its
28-dimensional adjoint
representation on the  $S^{7}_{1}$ and $S^{7}_{2}$ parts
of the antihermitian
matrix, plus the derivations $G_{2}$, since
\begin{equation}
Spin(0,8) = S^{7}_{1} \oplus S^{7}_{2} \oplus G_{2}
\end{equation}
\vspace{12pt}

On the hermitian Jordan algebra matrix $J_{3}^{{\bf{O}}}o$:
\begin{equation}
\left(
\begin{array}{ccc}
a & {\bf O}_{+} &  {\bf O}_{v} \\
& & \\
{\bf O}_{+}^\dagger & b &  {\bf O}_{-} \\
& & \\
{\bf O}_{v}^\dagger & {\bf O}_{-}^\dagger & -a-b
\end{array}
\right)
\end{equation}
the action of $Spin(0,8)$ (as a subgroup of $E_{6}$ and $F_{4}$)
leaves invariant each of the ${\bf O}_{v}$, ${\bf O}_{+}$, and
${\bf O}_{-}$ parts of the matrix, corresponding to the facts
that the gauge group $Spin(0,8)$:
\newline
\vspace{12pt}

does not act as a generalized supersymmetry to interchange
spacetime and fermions; and
\newline
\vspace{12pt}

does not interchange fermion particles and antiparticles.
\vspace{12pt}

\newpage

\section{3 Generations:  $E_{6}$, $E_{7}$, and $E_{8}$.}

In the $D_{4}-D_{5}-E_{6}$ model, the first generation
of spinor fermions is represented by the octonions ${\bf O}$,
the second by ${\bf O} \oplus {\bf O}$, and
the third by ${\bf O} \oplus {\bf O} \oplus {\bf O}$.
\newline
The global structure of the $D_{4}-D_{5}-E_{6}$ model
 with 8-dimensional spacetime and first generation fermions
is given by
\begin{equation}
E_{6} = F_{4} \oplus
\left(
\begin{array}{ccc}
a & {\bf O}_{+} &  {\bf O}_{v} \\
& & \\
{\bf O}_{+}^\dagger & b &  {\bf O}_{-} \\
& & \\
{\bf O}_{v}^\dagger & {\bf O}_{-}^\dagger & -a-b
\end{array}
\right)
\end{equation}
where
\begin{equation}
\left(
\begin{array}{ccc}
a & {\bf O}_{+} &  {\bf O}_{v} \\
& & \\
{\bf O}_{+}^\dagger & b &  {\bf O}_{-} \\
& & \\
{\bf O}_{v}^\dagger & {\bf O}_{-}^\dagger & -a-b
\end{array}
\right)
= J_{3}^{\bf O}o
\end{equation}
Here, the ${\bf O}_{+}$ and ${\bf O}_{}$ in
$J_{3}^{\bf O}o$ represent the ${\bf O}$ first generation
fermion particles and antiparticles.  Using the octonion basis
$\{ 1, e_{1}, e_{2}, e_{3}, e_{4}, e_{5}, e_{6}, e_{7} \}$ with
quaternionic subalgebra basis $\{ 1, e_{1}, e_{2}, e_{6} \}$ and
with octonion product Equation (4), the representation is:

\begin{equation}
\begin{array}{|c|c|} \hline
Octonion  & Fermion \: Particle \\
basis \: element & \\ \hline
1 & e-neutrino   \\ \hline
e_{1} & red \: up \: quark \\ \hline
e_{2} & green \: up \: quark \\ \hline
e_{6} & blue \: up \: quark  \\ \hline
e_{4} & electron \\ \hline
e_{3} & red \: down \: quark  \\ \hline
e_{5} & green \: down \: quark  \\ \hline
e_{7} & blue \: down \: quark  \\ \hline
\end{array}
\end{equation}

The ${\bf O}_{v}$ represents 8-dimensional spacetime.
\newline
\vspace{12pt}

To represent the ${\bf O} \oplus {\bf O}$ second generation
of fermions, you need a structure that generalizes Equation (52)
by having two copies of the fermion octonions.
\newline
\vspace{12pt}

The simplest such generalization is
\begin{equation}
?E? =
F_{4} \oplus
\left(
\begin{array}{ccc}
a & {\bf O}_{+} &  {\bf O}_{v} \\
& & \\
{\bf O}_{+}^\dagger & b &  {\bf O}_{-} \\
& & \\
{\bf O}_{v}^\dagger & {\bf O}_{-}^\dagger & -a-b
\end{array}
\right)
\oplus
\left(
\begin{array}{ccc}
0 & {\bf O}_{+} & 0 \\
& & \\
{\bf O}_{+}^\dagger & 0 &  {\bf O}_{-} \\
& & \\
0 & {\bf O}_{-}^\dagger & 0
\end{array}
\right)
\end{equation}
This proposal fails because the ${\bf O}_{+}$ and ${\bf O}_{}$
in $?E?$ are not embedded in  $J_{3}^{\bf O}o$ and therefore do
not transform like the first generation ${\bf O}_{+}$ and ${\bf O}_{}$
that are embedded in $J_{3}^{\bf O}o$.
\newline
\vspace{12pt}

Therefore, make the second simplest generalization:
\begin{equation}
??E?? =
F_{4} \oplus
\left(
\begin{array}{ccc}
a & {\bf O}_{+} &  {\bf O}_{v} \\
& & \\
{\bf O}_{+}^\dagger & b &  {\bf O}_{-} \\
& & \\
{\bf O}_{v}^\dagger & {\bf O}_{-}^\dagger & -a-b
\end{array}
\right)
\oplus
\left(
\begin{array}{ccc}
a & {\bf O}_{+} &  {\bf O}_{v} \\
& & \\
{\bf O}_{+}^\dagger & b &  {\bf O}_{-} \\
& & \\
{\bf O}_{v}^\dagger & {\bf O}_{-}^\dagger & -a-b
\end{array}
\right)
\end{equation}
The $??E??$ proposal also fails, because the algebraic structure
of the two copies of $J_{3}^{\bf O}o$ is incomplete.
\vspace{12pt}

To complete the algebraic structure ,
a third copy of $J_{3}^{\bf O}o$ must
be added, and all three copies must be related algebraically
like the imaginary quaternions $\{ i, j, k \}$.  This can be
done by tensoring $J_{3}^{\bf O}o$ with the imaginary quaternions
$S^{3} = SU(2) = Spin(3) = Sp(1)$.
\newline
\vspace{12pt}

Since the order of the octonions in ${\bf O} \oplus {\bf O}$
should be irrelevant (for example, the octonion pair
$\{ e_{i}, 1 \}$ should represent the same fermion as
the octonion pair  $\{ 1, e_{i} \}$), the structure must
include the derivation algebra of the automorphism group
of the quaternions, $SU(2)$.
\newline
\vspace{12pt}

The resulting structure is the 133-dimensional exceptional
Lie algebra $E_{7}$:
\begin{equation}
E_{7} =
F_{4} \oplus
SU(2)
\oplus
S^{3}
\otimes
\left(
\begin{array}{ccc}
a & {\bf O}_{+} &  {\bf O}_{v} \\
& & \\
{\bf O}_{+}^\dagger & b &  {\bf O}_{-} \\
& & \\
{\bf O}_{v}^\dagger & {\bf O}_{-}^\dagger & -a-b
\end{array}
\right)
\end{equation}
Therefore $E_{7}$ is the global structure algebra
of the second generation fermions.
\newline
\vspace{12pt}

For the third generation of fermions, note
that three algebraically independent copies of
$J_{3}^{\bf O}o$ generate seven copies, corresponding
to the imaginary octonions
$\{ 1, e_{1}, e_{2}, e_{3}, e_{4}, e_{5}, e_{6}, e_{7} \}$;
that the imaginary octonions can be represented by $S^{7}$;
and that the derivation algebra of the automorphism group
of the octonions is $G_{2}$.
\newline
\vspace{12pt}

Therefore, the 248-dimensional exceptional
Lie algebra $E_{8}$:
\begin{equation}
E_{8} =
F_{4} \oplus
G_{2}
\oplus
S^{7}
\otimes
\left(
\begin{array}{ccc}
a & {\bf O}_{+} &  {\bf O}_{v} \\
& & \\
{\bf O}_{+}^\dagger & b &  {\bf O}_{-} \\
& & \\
{\bf O}_{v}^\dagger & {\bf O}_{-}^\dagger & -a-b
\end{array}
\right)
\end{equation}
is the global structure algebra of the third
generation fermions.
\newline
\vspace{12pt}

\newpage

Since there are only three Lie algebras in the
series $E_{6}, E_{7}, E_{8}$, there are only
three generations of fermions.
\vspace{12pt}

The first-generation Lie algebra $E_{6}$ has
one copy of $J_{3}^{\bf O}o$, corresponding to
the complex imaginary $i$.
\vspace{12pt}

The second-generation Lie algebra $E_{7}$ adds
one more algebraically independent copy of $J_{3}^{\bf O}o$,
corresponding to the quaternionic imaginary $j$.
\newline
Together with the $i$ copy of $J_{3}^{\bf O}o$,
the $k$ copy is produced, so $E_{7}$ has in total
3 copies of $J_{3}^{\bf O}o$.
\vspace{12pt}

The third-generation Lie algebra $E_{8}$ adds
one more algebraically independent copy of $J_{3}^{\bf O}o$,
corresponding to the octonionic imaginary $e_{4}$.
\newline
Together with the $i=e_{1}, j=e_{2}, k=e_{6}$ copies of
$J_{3}^{\bf O}o$, the $e_{3}, e_{5}, e_{7}$ copies are
produced, so $E_{7}$ has in total 3 copies of $J_{3}^{\bf O}o$.
\vspace{12pt}

Therefore, transitions among generation of fermions
to a lower one involve elimination of algebraically independent
imaginaries.
\vspace{12pt}

{}From 2nd to 1st, it must effectively map $j \rightarrow 1$.
\newline
This transition can be done in one step.
\vspace{12pt}

{}From 3rd to 2nd, it must effectively map $e_{4} \rightarrow 1$.
This transition can also be done in one step.
\vspace{12pt}

{}From 3rd to 1st, it must effectively map $j,e_{4} \rightarrow 1$.
This transition cannot be done in one step.
\newline
The map $e_{4} \rightarrow 1$ only gets you to the 2nd generation.
\newline
You still need the additional map $j \rightarrow 1$
to get to the 1st.
\vspace{12pt}

Since the imaginaries $e_{4}$ and $j$ are orthogonal to
each other, there must be an intermediate step that is
effectively a phase shift of $\pi / 2$, and
the KOBAYASHI-MASKAWA PHASE angle parameter should be
\vspace{12pt}

\begin{equation}
\epsilon = \pi / 2
\end{equation}

\vspace{12pt}

\newpage

\section{Discrete Lattices and Dimensional Reduction.}

The physical manifolds of the $D_{4}-D_{5}-E_{6}$ model
should be representable in terms of discrete lattices
in order to be formulated as
a generalized Feynman checkerboard model.

\vspace{12pt}

General references on lattices, polytopes, and related
structures are the book of Conway and Sloane \cite{CON}
and the books \cite{COX1,COX2} and some papers \cite{COX3,COX4}
of Coxeter.

\vspace{12pt}

\newpage

\subsection{Discrete Lattice $D_{4}-D_{5}-E_{6}$ Model.}

As seen in section 2, 78-dimensional $E_{6}$ of
the $D_{4}-D_{5}-E_{6}$ model is made up of three parts:
\newline
38-dimensional space of antihermitian $3 \times 3$ octonion matrices;
\newline
14-dimensional space of the Lie algebra $G_{2}$; and
\newline
26-dimensional space of the Jordan algebra $J_{3}^{{\bf{O}}}o$.
\vspace{12pt}

\subsection{Antihermitian $3 \times 3$ Octonion Matrices.}
Antihermitian $3 \times 3$ octonion matrices:
\begin{equation}
\left(
\begin{array}{ccc}
S^{7}_{1} & {\bf O}_{+} &  {\bf O}_{v} \\
& & \\
-{\bf O}_{+}^\dagger & S^{7}_{2} &  {\bf O}_{-} \\
& & \\
-{\bf O}_{v}^\dagger & -{\bf O}_{-}^\dagger & -S^{7}_{1}-S^{7}_{2}
\end{array}
\right)
\end{equation}

\vspace{12pt}

Individually, ${\bf O}_{v}$, ${\bf O}_{+}$, and ${\bf O}_{-}$
can each be represented by
\newline
the 8-dimensional $E_{8}$ lattice,
as shown by Geoffrey Dixon \cite{DIX6}.
\newline
\vspace{12pt}

As Geoffrey Dixon \cite{DIX7} has shown, the two half-spinor
spaces ${\bf O}_{+}$, and ${\bf O}_{-}$ taken together can be
represented by the 16-dimensional Barnes-Wall lattice
$\Lambda_{16}$.
\vspace{12pt}

The Barnes-Wall lattices form a series of real dimension
$2^{n}$ for $n \geq 2$, and that the Barnes-Wall lattices
of real dimension 4 and 8 are the $D_{4}$ and $E_{8}$ lattices.
\vspace{12pt}

${\bf O}_{+}$, and ${\bf O}_{-}$ form the Shilov boundary
of a 16-complex dimensional bounded complex homogeneous
domain,
\newline
and Conway and Sloane \cite{CON} note that
\newline
 Quebbemann's 32-real dimensional lattice is a complex lattice
whose
\newline
16-real-dimensional real part is
the 16-real-dimensional Barnes-Wall $\Lambda_{16}$ lattice.
\vspace{12pt}

To represent all three ${\bf O}_{v}$, ${\bf O}_{+}$, and
${\bf O}_{-}$ together, it may be possible to use the
24-dimensional Leech lattice.
\vspace{12pt}

Such a Leech lattice representation is likely to be
related to the group
$$Spin(0,8) = S^{7} \Join S^{7} \Join G_{2}$$
where $\Join$ denotes the fibre product of the fibrations
$$Spin(7) \rightarrow Spin(8) \rightarrow S^{7}$$
and
$$G_{2} \rightarrow Spin(7) \rightarrow S^{7}$$
\vspace{12pt}

In a Leech lattice representation, it is likely that:
\vspace{12pt}

the ${\bf O}_{+}$ and ${\bf O}_{-}$ correspond
to the two $S^{7}$'s of the $Spin(0,8)$ fibrations,
and to the $X$-product of Martin Cederwall \cite{CED1}
and the $XY$-product on which Geoffrey Dixon is working; and
\vspace{12pt}

the ${\bf O}_{v}$ corresponds to a 7-dimensional
representation of the $G_{2}$ of the $Spin(0,8)$ fibrations.
\vspace{12pt}

The entire 26-dimensional space of traceless
antihermitian $3 \times 3$ octonion matrices may be
represented by the Lorentz Leech lattice $\Pi_{25,1}$,
\newline
which is closely related to the Monster group.
\vspace{12pt}

\newpage

\subsection{$S^{7} \oplus S^{7} \oplus G_{2} = D_{4}$.}
$S^{7}_{1} \oplus S^{7}_{2} \oplus G_{2} = D_{4}$ is
the Lie algebra of the $Spin(0,8)$ gauge group of
the $D_{4}-D_{5}-E_{6}$ model
(before spacetime dimensional reduction).
\vspace{12pt}

The gauge group $Spin(0,8)$ acts through its gauge bosons,
which can:
\newline
propagate through spacetime;
\newline
interact with fermion particles or antiparticles; and
\newline
interact with each other.
\vspace{12pt}

Propagation through spacetime is represented by the
8-dimensional vector representation of $Spin(0,8)$,
which in turn is represented by the octonions ${\bf O}_{v}$ and,
in discrete lattice version, by the $E_{8}$ lattice.
\vspace{12pt}

Interaction with fermion particles is represented by
the 8-dimensional +half-spinor representation of $Spin(0,8)$,
which in turn is represented by the octonions ${\bf O}_{+}$ and,
in discrete lattice version, individually by the $E_{8}$ lattice.
\vspace{12pt}

Interaction with fermion antiparticles is represented by
the 8-dimensional -half-spinor representation of $Spin(0,8)$,
which in turn is represented by the octonions ${\bf O}_{-}$ and,
in discrete lattice version, individually by the $E_{8}$ lattice.
\vspace{12pt}

Together, the two half-spinor representations are
\newline
represented by the Barnes-Wall $\Lambda_{16}$ lattice.
\vspace{12pt}

Interaction with other gauge bosons is represented by
the 28-dimensional adjoint representation of $Spin(0,8)$,
which in turn is represented by the exterior wedge bivector
product of two copies of the vector octonions ${\bf O}_{v}
\wedge {\bf O}_{v}$ and, in discrete lattice version,
by the exterior wedge bivector  product of two copies
of the $E_{8}$ lattice, $E_{8} \wedge E_{8}$.
\vspace{12pt}

\newpage

\subsection{Jordan algebra $J_{3}^{{\bf{O}}}o$.}
26-dimensional space of the Jordan algebra $J_{3}^{{\bf{O}}}o$:
\begin{equation}
\left(
\begin{array}{ccc}
a & {\bf O}_{+} &  {\bf O}_{v} \\
& & \\
{\bf O}_{+}^\dagger & b &  {\bf O}_{-} \\
& & \\
{\bf O}_{v}^\dagger & {\bf O}_{-}^\dagger & -a-b
\end{array}
\right)
\end{equation}
The 25+1 = 26-dimensional Lorentz Leech lattice $II_{25,1}$
can be used to represent the 26-dimensional Jordan
algebra $J_{3}^{{\bf{O}}}o$.
\vspace{12pt}

As seen in section 4.,the physically relevant group action
on $J_{3}^{{\bf{O}}}o$ for the Lagrangian dynamics of
the $D_{4}-D_{5}-E_{6}$ model is not action by
the global symmetry group $E_{6}$, but rather action
by the gauge group $Spin(0,8)$.
\vspace{12pt}

Since action on $J_{3}^{{\bf{O}}}o$ by the gauge group
$Spin(0,8)$ leaves invariant each of the
${\bf O}_{v}$, ${\bf O}_{+}$, and ${\bf O}_{-}$ parts of the matrix,
the only discrete lattice structure needed for the Lagrangian
dynamics of the $D_{4}-D_{5}-E_{6}$ model is the representation
of each of ${\bf O}_{v}$, ${\bf O}_{+}$, and ${\bf O}_{-}$ by
an $E_{8}$ lattice, and the full Lorentz Leech lattice $II_{25,1}$
is not needed.
\vspace{12pt}

\newpage

However, in case it may be useful to have a discrete lattice
description of global symmetries (for example, in looking at
generalized supersymmetric relationships among fermions and
bosons or at CPT symmetry), here are some characteristics of
the Lorentz Leech lattice $II_{25,1}$:

\vspace{12pt}

$II_{25,1}$ can be represented
(as in Conway and Sloane \cite{CON})
by the set of vectors
$\{ x_{0}, x_{1}, ... , x_{24} | x_{25} \}$
such that all the $x_{i}$ are in
${\bf Z}$ or all in ${\bf Z} + 1/2$
and satisfy
\newline
$x_{0} + ... + x_{24} - x_{25} \in 2{\bf Z}$

\vspace{12pt}

Let $w = (0,1,2,3, ... 23,24|70)$.
Since $0^{2}+1^{2}+2^{2}+...+24^{2} = 70^{2}$,
$w$ is an isotropic vector in $II_{25,1}$.

\vspace{12pt}

Then the Leech roots, or vectors $r$ in $II_{25,1}$ such
that $r \cdot r = 2$ and
$r \cdot w = -1$ are the vertices of
a 24-dimensional Leech lattice.

\vspace{12pt}

Also, $(w^{\bot} \cap II_{25,1})/w$ is a copy of the Leech lattice.

\vspace{12pt}

The 24-dimensional Leech lattice can be made up of 3 $E_{8}$
lattices, and so corresponds to the off-diagonal part of
$J_{3}^{{\bf{O}}}o$.

\vspace{12pt}

Each vertex of the Leech lattice has 196,560 nearest neighbors.

\vspace{12pt}

A space of 196,560+300+24 = 196,884 dimensions can be used
to represent the largest finite simple, group,
the Fischer-Greiss Monster of order
$$2^{46} \cdot 3^{20} \cdot 5^{9} \cdot 7^{6} \cdot 11^{2}
\cdot 13^{3} \cdot 17 \cdot 19
\cdot 23 \cdot 29 \cdot 31 \cdot 41 \cdot 47
\cdot 59 \cdot 71 =$$
$$= 808,017,424,794,512,875,886,459,904,$$
$$961,710,757,005,754,368,000,000,000$$

\newpage

\subsection{Lattice Dimensional Reduction.}

\subsubsection{HyperDiamond Lattices.}

The lattices of type $D_{n}$ are n-dimensional checkerboard
lattices, that is, the alternate vertices of a Zn hypercubic
lattice.  A general reference on lattices is Conway and
Sloane \cite{CON}.  For the n-dimensional HyperDiamond
construction from $D_{n}$, Conway and Sloane use an
n-dimensional glue vector $[1] = (0.5, ..., 0.5)$ (with
n $0.5$'s).
\vspace{12pt}

Consider $D_{3}$, the fcc close packing in 3-space.
Make a second $D_{3}$ shifted by the glue vector
$(0.5, 0.5, 0.5)$.
\vspace{12pt}

Then form the union $D_{3} \cup  ([1] + D_{3})$.
\vspace{12pt}

That is a 3-dimensional Diamond crystal.
\vspace{12pt}

When you do the same thing to get a 4-dimensional
HyperDiamond, you get $D_{8} \cup  ([1] + D_{8})$.
\newline
 The 4-dimensional HyperDiamond $D_{4} \cup  ([1] + D_{4})$
is the ${\bf{Z}}^{4}$ hypercubic lattice with null edges.
\vspace{12pt}

It is the lattice that Michael Gibbs \cite{GIB} uses
in his Ph. D. thesis advised by David Finkelstein.
\vspace{12pt}

When you construct an 8-dimensional HyperDiamond,
you get $D_{8} \cup  ([1] + D_{8})$ = $E_{8}$,
the fundamental lattice of the octonion structures in the
$D_{4}-D_{5}-E_{6}$ model described in
\href{http://xxx.lanl.gov/abs/hep-ph/9501252}{hep-ph/9501252}.
\vspace{12pt}

\subsubsection{Dimensional Reduction.}

Dimensional reduction of spacetime from 8 to 4 dimensions
takes the $E_{8}$ lattice into a $D_{4}$ lattice.
\newline
The $E_{8}$ lattice can be written in 7 different ways using
\newline
octonion coordinates with basis
\begin{equation}
\{ 1 ,e_{1},e_{2},e_{3},e_{4},e_{5},e_{6},e_{7} \}
\end{equation}
One way is:
\newline
16 vertices:
\begin{equation}
\pm 1, \pm e_{1}, \pm e_{2}, \pm e_{3}, \pm e_{4},
\pm e_{5}, \pm e_{6}, \pm e_{7}
\end{equation}
96 vertices:
\begin{equation}
\begin{array} {c}
(\pm 1 \pm e_{1} \pm e_{2} \pm e_{3}) / 2  \\
(\pm 1 \pm e_{2} \pm e_{5} \pm e_{7}) / 2  \\
(\pm 1 \pm e_{2} \pm e_{4} \pm e_{6}) / 2  \\
(\pm e_{4} \pm e_{5} \pm e_{6} \pm e_{7}) / 2  \\
(\pm e_{1} \pm e_{3} \pm e_{4} \pm e_{6}) / 2  \\
(\pm e_{1} \pm e_{3} \pm e_{5} \pm e_{7}) / 2  \\
\end{array}
\end{equation}
128 vertices:
\begin{equation}
\begin{array} {c}
(\pm 1 \pm e_{3} \pm e_{4} \pm e_{7}) / 2  \\
(\pm 1 \pm e_{1} \pm e_{5} \pm e_{6}) / 2  \\
(\pm 1 \pm e_{3} \pm e_{6} \pm e_{7}) / 2  \\
(\pm 1 \pm e_{1} \pm e_{4} \pm e_{7}) / 2  \\
(\pm e_{1} \pm e_{2} \pm e_{6} \pm e_{7}) / 2  \\
(\pm e_{2} \pm e_{3} \pm e_{4} \pm e_{7}) / 2  \\
(\pm e_{1} \pm e_{2} \pm e_{4} \pm e_{5}) / 2  \\
(\pm e_{2} \pm e_{3} \pm e_{5} \pm e_{6}) / 2  \\
\end{array}
\end{equation}

\vspace{12pt}

Consider the quaternionic subspace of the
octonions
\newline
with basis $\{ 1 ,e_{1},e_{2},e_{6} \}$ and
\newline
the $D_{4}$ lattice with origin nearest neighbors:
\newline
8 vertices:
\begin{equation}
\pm 1, \pm e_{1}, \pm e_{2}, \pm e_{6}
\end{equation}
and
\newline
16 vertices:
\begin{equation}
(\pm 1 \pm e_{1} \pm e_{2} \pm e_{6}) / 2  \\
\end{equation}

\vspace{12pt}

\newpage

Dimensional reduction of the $E_{8}$ lattice spacetime
to 4-dimensional spacetime reduces each of the $D_{8}$
lattices in the $E_{8}$ lattice to $D_{4}$ lattices.
\vspace{12pt}

Therefore, we should get a 4-dimensional HyperDiamond
$D_{4} \cup  ([1] + D_{4})$.
\vspace{12pt}

The 4-dimensional HyperDiamond $D_{4} \cup  ([1] + D_{4})$
is the ${\bf{Z}}^{4}$ hypercubic lattice with null edges.
\vspace{12pt}

It is the lattice that Michael Gibbs \cite{GIB} uses
in his Ph. D. thesis advised by David Finkelstein.
\vspace{12pt}

Here is an explicit construction of the 4-dimensional
HyperDiamond.
\vspace{12pt}

\newpage

START WITH THE 24 VERTICES OF A 24-CELL $D_{4}$:
\vspace{12pt}

\begin{equation}
\begin{array}{cccc}
+1   &    +1   &     0    &    0 \\
+1   &     0   &    +1    &    0 \\
+1   &     0   &     0    &   +1 \\
+1   &    -1   &     0    &    0 \\
+1   &     0   &    -1    &    0 \\
+1   &     0   &     0    &   -1 \\
-1   &    +1   &     0    &    0 \\
-1   &     0   &    +1    &    0 \\
-1   &     0   &     0    &   +1 \\
-1   &    -1   &     0    &    0 \\
-1   &     0   &    -1    &    0 \\
-1   &     0   &     0    &   -1 \\
 0   &    +1   &    +1    &    0 \\
 0   &    +1   &     0    &   +1 \\
 0   &    +1   &    -1    &    0 \\
 0   &    +1   &     0    &   -1 \\
 0   &    -1   &    +1    &    0 \\
 0   &    -1   &     0    &   +1 \\
 0   &    -1   &    -1    &    0 \\
 0   &    -1   &     0    &   -1 \\
 0   &     0   &    +1    &   +1 \\
 0   &     0   &    +1    &   -1 \\
 0   &     0   &    -1    &   +1 \\
 0   &     0   &    -1    &   -1 \\
\end{array}
\end{equation}

\vspace{12pt}

\newpage

SHIFT THE LATTICE BY A GLUE VECTOR,
\newline
BY ADDING
\vspace{12pt}

\begin{equation}
\begin{array}{cccc}
  0.5   &     0.5   &     0.5   &     0.5 \\
\end{array}
\end{equation}

\vspace{12pt}

TO GET 24 MORE VERTICES $[1] + D_{4}$:
\vspace{12pt}

\begin{equation}
\begin{array}{cccc}
+1.5   &    +1.5   &     0.5   &     0.5 \\
+1.5   &     0.5   &    +1.5   &     0.5 \\
+1.5   &     0.5   &     0.5   &    +1.5 \\
+1.5   &    -0.5   &     0.5   &     0.5 \\
+1.5   &     0.5   &    -0.5   &     0.5 \\
+1.5   &     0.5   &     0.5   &    -0.5 \\
-0.5   &    +1.5   &     0.5   &     0.5 \\
-0.5   &     0.5   &    +1.5   &     0.5 \\
-0.5   &     0.5   &     0.5   &    +1.5 \\
-0.5   &    -0.5   &     0.5   &     0.5 \\
-0.5   &     0.5   &    -0.5   &     0.5 \\
-0.5   &     0.5   &     0.5   &    -0.5 \\
 0.5   &    +1.5   &    +1.5   &     0.5 \\
 0.5   &    +1.5   &     0.5   &    +1.5 \\
 0.5   &    +1.5   &    -0.5   &     0.5 \\
 0.5   &    +1.5   &     0.5   &    -0.5 \\
 0.5   &    -0.5   &    +1.5   &     0.5 \\
 0.5   &    -0.5   &     0.5   &    +1.5 \\
 0.5   &    -0.5   &    -0.5   &     0.5 \\
 0.5   &    -0.5   &     0.5   &    -0.5 \\
 0.5   &     0.5   &    +1.5   &    +1.5 \\
 0.5   &     0.5   &    +1.5   &    -0.5 \\
 0.5   &     0.5   &    -0.5   &    +1.5 \\
 0.5   &     0.5   &    -0.5   &    -0.5 \\
\end{array}
\end{equation}

\vspace{12pt}

\newpage

FOR THE NEW COMBINED LATTICE $D_{4} \cup ([1] + D_{4})$,
\newline
THESE ARE 6 OF THE NEAREST NEIGHBORS
\newline
TO THE ORIGIN:
\vspace{12pt}

\begin{equation}
\begin{array}{cccc}
-0.5   &    -0.5   &     0.5   &     0.5 \\
-0.5   &     0.5   &    -0.5   &     0.5 \\
-0.5   &     0.5   &     0.5   &    -0.5 \\
 0.5   &    -0.5   &    -0.5   &     0.5 \\
 0.5   &    -0.5   &     0.5   &    -0.5 \\
 0.5   &     0.5   &    -0.5   &    -0.5 \\
\end{array}
\end{equation}

\vspace{12pt}

HERE ARE 2 MORE THAT COME FROM
\newline
ADDING THE GLUE VECTOR TO LATTICE VECTORS
\newline
THAT ARE NOT NEAREST NEIGHBORS OF THE ORIGIN:
\vspace{12pt}

\begin{equation}
\begin{array}{cccc}

 0.5   &     0.5   &     0.5   &     0.5 \\
-0.5   &    -0.5   &    -0.5   &    -0.5 \\
\end{array}
\end{equation}

\vspace{12pt}

THEY COME, RESPECTIVELY, FROM ADDING
\newline
THE GLUE VECTOR TO:
\vspace{12pt}

THE ORIGIN
\vspace{12pt}

\begin{equation}
\begin{array}{cccc}
  0   &     0   &     0   &     0 \\
\end{array}
\end{equation}

\vspace{12pt}

ITSELF;
\vspace{12pt}

AND
\vspace{12pt}

\newpage

THE LATTICE POINT
\vspace{12pt}

\begin{equation}
\begin{array}{cccc}
  -1   &     -1   &     -1   &     -1 \\
\end{array}
\end{equation}

\vspace{12pt}

WHICH IS SECOND ORDER, FROM
\vspace{12pt}

\begin{equation}
\begin{array}{cccc}
  -1   &     -1   &     0   &     0 \\
plus &&&  \\
  0   &     0   &     0   &     0 \\
\end{array}
\end{equation}

\vspace{12pt}

FROM
\vspace{12pt}

\begin{equation}
\begin{array}{cccc}
  -1   &     0   &     -1   &     0 \\
plus &&&  \\
  0   &     -1   &     0   &     -1 \\
\end{array}
\end{equation}

\vspace{12pt}

OR FROM
\vspace{12pt}

\begin{equation}
\begin{array}{cccc}
  -1   &     0   &     0   &     -1 \\
plus &&&  \\
  0   &     -1   &     -1   &     0 \\
\end{array}
\end{equation}

\vspace{12pt}

\newpage

That the $E_{8}$ lattice is, in a sense,
fundamentally 4-dimensional
can be seen from several points of view:
\newline
\vspace{12pt}

the $E_{8}$ lattice nearest neighbor vertices have
only 4 non-zero coordinates,
like 4-dimensional spacetime with speed of light
$c$ = $\sqrt{3}$,
rather than 8 non-zero coordinates,
like 8-dimensional spacetime with speed of light $c$ = $\sqrt{7}$,
so the $E_{8}$ lattice light-cone structure appears to be
4-dimensional rather than 8-dimensional;
\newline
\vspace{12pt}

the representation of the $E_{8}$ lattice by quaternionic icosians,
as described by Conway and Sloane \cite{CON};
\newline
\vspace{12pt}

the Golden ratio construction of the $E_{8}$ lattice from
the $D_{4}$ lattice, which has a 24-cell nearest neighbor polytope
(The construction starts with the 24 vertices of a 24-cell,
then adds Golden ratio points on each of the 96 edges of the 24-cell,
then extends the space to 8 dimensions by considering
the algebraicaly independent $\sqrt{5}$ part of the coordinates to
be geometrically independent, and
finally doubling the resulting 120 vertices in 8-dimensional
space (by considering both the $D_{4}$ lattice and its dual $D_{4}^{\ast}$)
to get the 240 vertices of the $E_{8}$ lattice nearest neighbor
polytope (the Witting polytope); and
\newline
\vspace{12pt}

the fact that the 240-vertex Witting polytope,
the $E_{8}$ lattice nearest neighbor polytope,
most naturally lives in 4 complex dimensions,
where it is self-dual, rather than in 8 real dimensions.
\newline
\vspace{12pt}

Some more material on such things can be found at
\newline
\href{http://www.gatech.edu/tsmith/home.html}{WWW
URL http://www.gatech.edu/tsmith/home.html} \cite{SMI6}.

\vspace{12pt}

In referring to Conway and Sloane \cite{CON}, bear in mind that
they use the convention (usual in working with lattices) that
the norm of a lattice distance is the square of the length of
the lattice distance.
\newline
\vspace{12pt}

It is also noteworthy that the number of vertices
in shells of an $E_{8}$ lattice increase monotonically
as the radius of the shell increases,
\newline
while cyclic relationships (see Conway and Sloane \cite{CON})
appear in the number of vertices in shells of a $D_{4}$ lattice.
\vspace{12pt}

\newpage

\subsection{Discrete Lattice Effects of $8 \rightarrow 4 \; dim$.}

$Spin(0,8)$ acts on the octonions ${\bf O}$, the lattice version
of which is the $E_{8}$ lattice.

\vspace{12pt}

Each vertex in the $E_{8}$ lattice has 240 nearest neighbors,
the inner shell of the $E_{8}$ lattice.

\vspace{12pt}

Geoffrey Dixon \cite{DIX7} shows that the 240 vertices
in the $E_{8}$ inner shell break down with respect to
the two 4-dimensional subspaces of ${\bf O}$, each
represented by the inner shell of a $D_{4}$ lattice, as

\begin{equation}
\begin{array}{ccc}
<U,0>        &          \rightarrow & 24 \; elements \\
<0,V>        &          \rightarrow & 24 \; elements  \\
<W,X> (WX \ast = +/- qm) &  \rightarrow & 192 \; elements
\end{array}
\end{equation}

where $D_{4}^{\ast}$ is the dual lattice to the $D_{4}$ lattice, and
where $U,V \in D_{4}$, $W \in D_{4}^{\ast}$, and
$X \in \{ \pm 1, \pm i, \pm j, \pm k \} \subset D_{4}^{ \ast }$

\vspace{12pt}

The two 24-element sets each have the group structure of the
binary tetrahedral group, also the group of 24 quaternion units,
and the 24 elements would represent the root vectors of the
$Spin0,(8)$ $D_{4}$ Lie algebra in the 4-dimensional
space of the $D_{4}$ lattice.

\vspace{12pt}

The 192 element set is the Weyl group of the
$Spin0,(8)$ $D_{4}$ Lie algebra.
\newline
The Weyl group is the group of reflections in
the hyperplanes (in the $D_{4}$ 4-dimensional space)
that are orthogonal to the 24 root vectors.

\vspace{12pt}

If the 8-dimensional $E_{8}$ spacetime is reduced
to the 4-dimensional $D_{4}$ spacetime, then

\begin{equation}
\begin{array}{ccc}
<U,0> & \rightarrow & 24 \; elements \; of \; D_{4} \; lattice \; \\
&&inner \; shell\\
<0,V> & \rightarrow & 24 \; elements \; of \; binary \; \\
&&tetrahedral \; group\\
<W,X> (WX \ast = +/- qm) &  \rightarrow & 192 \;
elements \; of \; Weyl \; group \; \\
&&of \; reduced \; gauge \; group
\end{array}
\end{equation}

where $U,V \in D_{4}$, $W \in D_{4}^{\ast}$, and
$X \in \{\pm 1, \pm i, \pm j, \pm k \} \subset D_{4}^{\ast}$

\vspace{12pt}

The 24-element $D_{4}$ lattice inner shell formed by
the $U$ elements of Equation (76) form the second shell
of the reduced 4-dimensional spacetime $D_{4}$ lattice,
as described in the preceding Subsection 7.5.
\newline
Denote this set by $24U$.

\vspace{12pt}

The 24-element $D_{4}$ lattice inner shell formed by
the $V$ elements of Equation (76) form the 24-element
finite group (binary tetrahedral group of unit quaternions)
that is the Weyl group of the internal symmetry gauge group of
the 4-dimensional $D_{4}-D_{5}-E_{6}$ model.
\newline
Denote this group by $24V$.

\vspace{12pt}

The 192-element Weyl group of the $D_{4}$ Lie algebra of
$Spin(0,8)$ is made up of pairs, the first of which is
an element of the 24-element set of $W$ elements.
\newline
Denote that 24-element set by $24W$.
\newline
The second part of a pair is an element of the 8-element
set of $X$ elements.
\newline
Denote that 8-element set by $8X$.
\newline
\vspace{12pt}

After dimensional reduction of spacetime, $Spin(0,8)$ is
too big to act as an isotropy group on spacetime, as it acted
in 8-dimensional spacetime, which is of the form
$$Spin(2,8) / Spin(2,0) \times Spin(0,8)$$.

\vspace{12pt}

Therefore, in 4-dimensional spacetime, the 28 infinitesimal
generators of $Spin(0,8)$ (which act as 28 gauge bosons in
8-dimensional $D_{4}-D_{5}-E_{6}$ physics) cannot interact
according to the commutation relations of $Spin(0,8)$, but
must interact according to commutation relations of smaller
groups that can act on 4-dimensional spacetime.

\vspace{12pt}

What type of action should these smaller groups have on
4-dimensional spacetime?

\vspace{12pt}

Isotropy action is sufficient in the case of the 28
$Spin(0,8)$ infinitesimal generators in the 8-dimensional
theory, because the gauge boson part of the Lagrangian
$\int_{8-dim} F \wedge \star F$ is an integral over
8-dimensional spacetime using a uniform measure that is
the same for all 28 gauge boson infinitesimal generators.
Physically, the force strength of the $Spin(0,8)$ gauge
group can be taken to be $1$ because there is only one
gauge group in the 8-dimensional $D_{4}-D_{5}-E_{6}$ model.
\newline
However, as we shall see now, isotropy action is not
sufficient for the gauge groups after dimensional reduction.

\vspace{12pt}

After reduction to 4-dimensional spacetime, the 28
infinitesimal generators will have to regroup into
more that one smaller groups, each of which will have
its own force strength.  The 4-dimensional Lagrangian
will be the sum of more than one Lagrangians of the
form $\int_{4-dim} F \wedge \star F$, each of which
will use a different measure in integrating over
4-dimensional spacetime.
\newline
A factor in determining the relative strengths of
the 4-dimensional forces will be the relative magnitude
of the measures over 4-dimensional spacetime.
Therefore, the measure information should be carried
in the $F$ of the 4-dimensional Lagrangian
$\int_{4-dim} F \wedge \star F$, which is different
for each force, rather than in the overall $\int_{4-dim}$,
which should be uniform for all terms in the total
4-dimensional Lagrangian of the $D_{4}-D_{5}-E_{6}$ model.
\newline
\vspace{12pt}

In the 4-dimensional $D_{4}-D_{5}-E_{6}$ model, the
small gauge groups must act transitively on
4-dimensional spacetime, so that they can carry
the measure information.  Physically, the gauge bosons
of the different gauge groups see spacetime differently
\newline
(see
\href{http://www.gatech.edu/tsmith/See.html}{WWW
URL http://www.gatech.edu/tsmith/See.html}
\cite{SMI6} ).

\vspace{12pt}

Since the 4-dimensional $D_{4}-D_{5}-E_{6}$ model gauge groups
\newline
are not isotropy groups of 4-dimensional spacetime,
\newline
but actually act transitively on 4-dimensional manifolds,
\newline
they are not exactly conventional "local symmetry" gauge groups.
\newline
In the conventional "local symmetry" picture, you can
put a gauge boson infinitesimal generator $x(G)$ of
the gauge group $G$ at each point $p$ of
the spacetime base manifold, with the choice of $x(G)$ made
independently at each point $p$.
\newline
In the $D_{4}-D_{5}-E_{6}$ model picture, the choice of $x(G)$
at a point not only is the choice of a gauge boson, but also
of a "translation" direction in the 4-dimensional spacetime.
\newline
\vspace{12pt}

When you choose a gauge boson, say a 'red-antiblue gluon"
\newline
of the $SU(3)$ color force gauge group, then
\newline
does your "choice" of "translation direction" fix a
\newline
physical direction of propagation (say, $(+1-i-j-k)/2$
\newline
in quaternionic coordinates for the future lightcone) ?
\newline
\vspace{12pt}

If it does, the model doesn't work right.
\newline
Fortunately for the $D_{4}-D_{5}-E_{6}$ model, there is
\newline
one more choice to be made independently at each point, and
\newline
that is the choice to put any given element of the gauge group $G$
\newline
in corresponce with with any element of the isotropy subgroup $K$
\newline
or with any direction in the 4-dimensional manifold $G / K$
\newline
on which $G$ acts transitively.

\vspace{12pt}

Therefore, at any given point in the 4-dimensional spacetime,
\newline
you can choose the "red-antiblue gluon" gauge boson
\newline
(or any other gauge boson) and
\newline
the $(+1-i-j-k)/2$ direction (or any other direction),
\newline
and independent choices can be made at all points in
\newline
the 4-dimensional spacetime of the $D_{4}-D_{5}-E_{6}$ model.

\vspace{12pt}

Since choices of gauge boson and
\newline
direction of propagation are both made at once,
\newline
it is natural in the lattice $D_{4}-D_{5}-E_{6}$ model to picture
\newline
the gauge bosons as living on the links of the spacetime lattice,
\newline
with the fermion particles and antiparticles living on the vertices.

\vspace{12pt}

This type of structure might not work consistently in
a model with less symmetry than the $D_{4}-D_{5}-E_{6}$ model.
\newline
Particularly, the triality symmetry of 8-dimensional spacetime
\newline
with the 8-dimensional half-spinor representation spaces of
\newline
the first-generation fermion particles and antiparticles
\newline
means that the gauge boson symmetry group,
\newline
which must act on fermion particles and
antiparticles in a natural way,
\newline
also acts transitively on spacetime in a natural way.
\newline
For a discussion of what types of "generalized supersymmetry"
\newline
symmetries are useful, and why the $D_{4}-D_{5}-E_{6}$ model
\newline
probably has the most useful symmetries of any model, see
\newline
\href{http://xxx.lanl.gov/abs/hep-th/9306011}{hep-th/9306011}
\cite{SMI3} ).

\vspace{12pt}

Even after dimensional reduction of spacetime, there is a
residual symmetry relationship between the fermion
representation spaces and spacetime.
\newline
Perhaps that residual symmetry might be a way to relate
the results of the $D_{4}-D_{5}-E_{6}$ model to the results
of the 4-dimensional lattice model of Finkelstein and Gibbs.
\cite{GIB}

\vspace{12pt}

What can these smaller groups be?  Since 4-dimensional
spacetime has quaternionic structure, they must act
transitively on 4-dimensional manifolds with
quaternionic structure.
\newline
Such manifolds have been classified by Wolf \cite{WOL}, and
they are

\begin{equation}
\begin{array}{|c|c|c|}
\hline
M & Symmetric \: Space   & Gauge \: Group  \\
\hline
& & \\
S^{4} & Spin(5) \over Spin(4) & Spin(5) \\
& & \\
{\bf C}P^2 & SU(3) \over {SU(2) \times U(1)}
& SU(3) \\
& & \\
S^2 \times S^2 & 2 \: copies \: of \; \left( SU(2) \over U(1) \right)
& SU(2) \\
& & \\
S^1 \times S^1 \times S^1
\times S^1 & 4 \: copies \: of \; U(1) & U(1)  \\
& & \\
\hline
\end{array}
\end{equation}

Therefore the 4 forces have gauge groups
\newline
$Spin(5)$ (10 infinitesimal generators)
\newline
$SU(3)$ (8 infinitesimal generators)
\newline
$SU(2)$ (2 copies, each with 3 infinitesimal generators)
\newline
and
\newline
$U(1)$ (4 copies, each with 1 infinitesimal generator)
\newline
that account for all 28 of the $Spin(0,8)$ gauge bosons.

\vspace{12pt}

There are two cases in which 4-dimensional spacetime is
made up of multiple copies of lower-dimensional manifolds.

\vspace{12pt}

Two copies of the gauge group $SU(2)$ act on 2 copies of $S^{2}$.
\newline
Each $SU(2)$ has 3 infinitesimal generators,
\newline
the three
weak bosons $\{ W_{-}, W_{0}, W_{+}$.
\newline
Since a given weak boson cannot carry a different charge in
\newline
different parts of 4-dimensional spacetime,
\newline
the 2 copies of $SU(2)$ must be aligned consistently.
\newline
This means that there is physically only one $SU(2)$ weak force
gauge group,
\newline
and that there are 3 degrees of freedom due to
\newline
3 $Spin(0,8)$ infinitesimal generators that are not used.
\newline
Denote them by $3-SU(2)$.

\vspace{12pt}

Four copies of the gauge group $U(1)$ act on 4 copies of $S^{1}$.
\newline
Each $U(1)$ has 1 infinitesimal generator, the photon.
\newline
Since a given photon should be the same in all parts
of 4-dimensional spacetime,
\newline
the 4 copies of $U(1)$ must be aligned consistently.
\newline
This means that there is physically only one $U(1)$
electromagnetic gauge group,
\newline
and that there are 3 degrees of freedom due to
\newline
3 $Spin(0,8)$ infinitesimal generators that are not used.
\newline
Denote them by $3-U(1)$.

\vspace{12pt}

How does all this fit into the structures
$24U$, $24V$, $24W$, and $8X$ ?

\vspace{12pt}

$24U$ and $24W$ are the $D_{4}$ and $D_{4}^{\ast}$
of the 4-dimensional lattice spacetime as described
in the preceding Subsection 7.5  .

\vspace{12pt}

The $8X$, being part of a 192-element product with $24W$,
should represent a gauge group that is closely connected
to spacetime.  $8X$ represents the 8-element Weyl group
$S_{2}^{3}$ of the gauge group $Spin(5)$.
\newline
By the MacDowell-Mansouri mechanism \cite{MAC},
the $Spin(5)$ gauge group accounts for Einstein-Hilbert gravity.

\vspace{12pt}

$24V$, being entirely from the part of 8-dimensional
spacetime that did not survive dimensional reduction,
should represent the Weyl groups of gauge groups of
internal symmetries.
\newline
As $24V$ is the 24-element binary tetrahedral group of
unit quaternions, the inner shell of the $D_{4}$ lattice,
it has a 4-element subgroup $S_{2}^{2}$ made up of unit
complex numbers, the inner shell of the Gaussian lattice.
\newline
$24V / S_{2}^{2}$ is the 6-element group $S_{3}$, which
is the Weyl group of the $SU(3)$ gauge group of the color force.
\newline
The $S_{2}^{2}$ is 2 copies of the Weyl group of the
$SU(2)$ gauge group of the weak force.
\newline
As $U(1)$ is Abelian, and so has the identity for its Weyl
group, 4 copies of the $U(1)$ gauge group of electromagnetism
can be said to be included in the gauge groups of which
$24V$ is the Weyl group.

\vspace{12pt}

Therefore, $24V$ gives us the gauge groups of the Standard Model,
$SU(3) \times SU(2) \times U(1)$,
\newline
plus the extra 6 degrees of freedom $3-SU(2)$ and $3-U(1)$
that were discussed above.

\vspace{12pt}

It will be seen in Section 8.2 that 5 of the 6 extra degrees of
freedom are a link between the Higgs sector of the Standard Model
and conformal symmetry related to gravity, and the 6th, a copy
of U(1), accounts for the complex phase of propagators in the
$D_{4}-D_{5}-E_{6}$ model.

\vspace{12pt}

{}From the continuum limit viewpoint of Section 8.2,
\newline
those 6 degrees of freedom are combined with the gravity sector
\newline
to expand the 10-dimensional de Sitter $Spin(5)$ group
\newline
of the MacDowell-Mansouri mechanism
\newline
to the 15-dimensional conformal $Spin(2,4)$ group
\newline
plus the $U(1)$ of the complex propagator phase,
\newline
which in turn can all be combined into
\newline
one copy of 16-dimensional $U(4)$.

\vspace{12pt}

{}From the discrete Weyl group point of view of this section,
the $3-SU(2)$ and $3-U(1)$ degrees of freedom are represented
by the Weyl group $S_{2}$.
\newline
The $S_{2}$ can be identified with the 3 $S_{2}$'s of the
gravity $Spin(5)$ $S_{2}^{3}$ in 3 ways, so the identification
effectively expands the gravity sector Weyl group by a factor of 3
from $S_{2}^{3} = S_{2}^{2} \times S_{2}$ to
$S_{2}^{2} \times S_{3}$, which is the Weyl group of
the conformal group $Spin(2,4)$ and also the Weyl group of $U(4)$.

\vspace{12pt}

\newpage

\section{Continuum Limit Effects of $8 \rightarrow 4 \; dim$.}

The dimensional reduction breaks the gauge group $Spin(0,8)$
into gravity plus the Standard Model.
\newline
\vspace{12pt}

The following sections discuss the effects of dimensional reduction
on the terms of the 8-dimensional Lagrangian
\begin{equation}
\int_{V_{8}} F_{8} \wedge \star F_{8} + \partial_{8}^{2}
\overline{\Phi_{8}} \wedge \star \partial_{8}^{2} \Phi_{8} +
\overline{S_{8\pm}} \not \! \partial_{8} S_{8\pm}
\end{equation}
of the $D_{4}-D_{5}-E_{6}$ model discussed in Section 5.
\newline
\vspace{12pt}

and the phenomenological results of calculations
\newline
based on the 4-dimensional structures.

\newpage

\subsection{Scalar part of the Lagrangian}

The scalar part of the 8-dimensional Lagrangian is

$$
\int_{V_{8}}
\partial_{8}^{2} \overline{\Phi_{8}} \wedge \star
\partial_{8}^{2} \Phi_{8}
$$

As shown in chapter 4 of G\"{o}ckeler and Sch\"{u}cker
\cite{GOC}, $\partial_{8}^{2} \Phi_{8}$ can be represented
as an 8-dimensional curvature $F_{H8}$, giving

$$
\int_{V_{8}} F_{H8} \wedge \star F_{H8}
$$

When spacetime is reduced to 4 dimensions, denote the
surviving 4 dimensions by $4$ and the reduced
4 dimensions by $\perp 4$.

\vspace{12pt}

Then, $F_{H8} = F_{H44} + F_{H4\perp 4} + F_{H\perp 4 \perp 4}$,
where

$F_{H44}$ is the part of $F_{H8}$ entirely in the
surviving spacetime;

$F_{H4 \perp 4}$ is the part of $F_{H8}$ partly in the
surviving spacetime and partly in the reduced spacetime; and

$F_{H\perp 4 \perp 4}$ is the part of $F_{H8}$ entirely in the
reduced spacetime;

\vspace{12pt}

The 4-dimensional Higgs Lagrangian is then:

$\int (F_{H44} + F_{H4 \perp 4} + F_{H\perp 4 \perp 4})
\wedge
\star (F_{H44} + F_{H4 \perp 4} + F_{H\perp 4 \perp 4}) =$

$=\int (F_{H44} \wedge \star F_{H44} +
F_{H4 \perp 4} \wedge \star F_{H4 \perp 4} +
F_{H\perp 4 \perp 4} \wedge \star F_{H\perp 4 \perp 4})$.

\vspace{12pt}

As all possible paths should be taken into account in
the sum over histories path integral picture of quantum
field theory, the terms involving the reduced 4 dimensions,
$\perp 4$, should be integrated over the reduced 4 dimensions.

\vspace{12pt}

Integrating over the reduced 4 dimensions, $\perp 4$, gives

$\int \left(  F_{H44} \wedge \star F_{H44}  +
 \int_{\perp 4} F_{H4 \perp 4} \wedge \star F_{H4 \perp 4} +
\int_{\perp 4} F_{H\perp 4 \perp 4} \wedge
\star F_{H\perp 4 \perp 4} \right)$.

\vspace{12pt}

\subsubsection{First term $ F_{H44} \wedge \star F_{H44}$}

\vspace{12pt}

The first term is just $\int F_{H44} \wedge \star F_{H44}$.

Since they are both $SU(2)$ gauge boson terms, this term,
in 4-dimensional spacetime, just merges into the $SU(2)$ weak
force term $\int F_{w} \wedge \star F_{w}$.

\vspace{12pt}

\subsubsection{Third term $\int_{\perp 4} F_{H\perp 4 \perp 4}
\wedge \star F_{H\perp 4 \perp 4}$}

\vspace{12pt}

The third term,
$ \int \int_{\perp 4} F_{H\perp 4 \perp 4} \wedge
\star F_{H\perp 4 \perp 4}$,
after integration over $\perp 4$,
produces terms of the form

$ \lambda (\overline{\Phi} \Phi)^{2} - \mu^{2}
\overline{\Phi} \Phi$
by a process similar to the Mayer mechanism
(see Mayer's paper \cite{MAY} for a description of
the Mayer mechanism, a geometric Higgs mechanism).

\vspace{12pt}

The Mayer mechanism is based on Proposition 11.4 of

chapter 11 of volume I of Kobayashi and Nomizu \cite{KOB},
stating that:

$2 F_{H\perp 4 \perp 4}(X,Y) = [\Lambda(X), \Lambda(Y)] -
\Lambda([X,Y])$,

where $\Lambda$ takes values in the $SU(2)$ Lie algebra.

\vspace{12pt}

If the action of the Hodge dual $\star$ on $\Lambda$ is
such that

$\star \Lambda = - \Lambda$ and $\star [\Lambda, \Lambda] =
[\Lambda, \Lambda]$,

then

$F_{H\perp 4 \perp 4}(X,Y) \wedge
\star F_{H\perp 4 \perp 4}(X,Y) =
(1/4)([\Lambda(X), \Lambda(Y)]^{2} - \Lambda([X,Y])^{2} )$.

\vspace{12pt}

If integration of $\Lambda$ over $\perp 4$ is
$\int_{\perp 4} \Lambda \propto \Phi = (\Phi^{+}, \Phi^{0})$,
then

\vspace{12pt}

$\int_{\perp 4} F_{H\perp 4 \perp 4} \wedge
\star F_{H\perp 4 \perp 4} = $
$ (1/4) \int_{\perp 4} [\Lambda(X),\Lambda(Y)]^{2} -
\Lambda([X,Y])^{2} = $

\vspace{12pt}

$= (1/4) [ \lambda ( \overline{\Phi} \Phi)^{2} - \mu^{2}
\overline{\Phi} \Phi ]$,

\vspace{12pt}

where $\lambda$ is the strength of the scalar field
self-interaction,
$\mu^{2}$ is the other constant in the Higgs potential, and
where $\Phi$ is a 0-form taking values in
the $SU(2)$ Lie algebra.

\vspace{12pt}

The $SU(2)$ values of $\Phi$ are represented by complex

$SU(2) = Spin(3)$ doublets $\Phi = (\Phi^{+}, \Phi^{0})$.

\vspace{12pt}

In real terms,
 $\Phi^{+} = (\Phi_{1} + i \Phi_{2})/ \sqrt{2}$
and
$\Phi^{0} = (\Phi_{3} + i \Phi_{4})/ \sqrt{2}$,

so $\Phi$ has 4 real degrees of freedom.

\vspace{12pt}

In terms of real components,
$\overline{\Phi} \Phi = (\Phi_{1}^{2} + \Phi_{2}^{2} +
\Phi_{3}^{2} + \Phi_{4}^{2})/2 $.

\vspace{12pt}

The nonzero vacuum expectation value of the

$ \lambda (\overline{\Phi} \Phi)^{2} - \mu^{2}
\overline{\Phi} \Phi$ term
is $v = \mu / \sqrt{\lambda}$, and

$<\Phi^{0}> = <\Phi_{3}> = v / \sqrt{2}$.

\vspace{12pt}

In the unitary gauge, $\Phi_{1} = \Phi_{2} = \Phi_{4} = 0$,

and

$\Phi = (\Phi^{+}, \Phi^{0}) = (1/ \sqrt{2})(\Phi_{1} + i
\Phi_{2}, \Phi_{3} + i \Phi_{4}) = (1/ \sqrt{2})
(0, v + H)$,

\vspace{12pt}

where $\Phi_{3} = (v + H) / \sqrt{2}$,

$v$ is the Higgs potential vacuum expectation value, and

$H$ is the real surviving Higgs scalar field.

\vspace{12pt}

Since $\lambda = \mu^{2} / v^{2}$ and $\Phi = (v + H)
/ \sqrt{2}$,

\vspace{12pt}

$(1/4)[ \lambda (\overline{\Phi} \Phi)^{2} - \mu^{2}
\overline{\Phi} \Phi ] = $

\vspace{12pt}

$= (1/16) (\mu^{2} / v^{2})(v + H)^{4} -
(1/8) \mu^{2} (v + H)^{2} = $

\vspace{12pt}

$= (1/16) [ \mu^{2} v^{2} + 4 \mu^{2} vH +
6 \mu^{2} H^{2} + 4 \mu^{2} H^{3} / v + \mu^{2} H^{4} /
v^{2} - 2 \mu^{2} v^{2} - $

$- 4 \mu^{2} v H - 2 \mu^{2} H^{2} ] = $

\vspace{12pt}

$= (1/4) \mu^{2} H^{2} - (1/16) \mu^{2} v^{2}
[ 1 - 4 H^{3} / v^{3} - H^{4} / v^{4} ] $.

\vspace{12pt}

\subsubsection{Second term $F_{H4 \perp 4} \wedge
\star F_{H4 \perp 4}$}

\vspace{12pt}

The second term,

$\int_{\perp 4} F_{H4 \perp 4} \wedge \star F_{H4 \perp 4}$,

gives $\int \partial \overline{\Phi} \partial \Phi$,
by a process similar to the Mayer mechanism
(see Mayer's paper \cite{MAY} for a description of
the Mayer mechanism, a geometric Higgs mechanism).

\vspace{12pt}

{}From Proposition 11.4 of chapter 11 of volume I of Kobayashi
and Nomizu \cite{KOB}:

$2 F_{H4 \perp 4}(X,Y) = [\Lambda(X), \Lambda(Y)] -
\Lambda([X,Y])$,

where $\Lambda$ takes values in the $SU(2)$ Lie algebra.

\vspace{12pt}

For example, if the $X$ component of $F_{H4 \perp 4}(X,Y)$
is in the surviving $4$ spacetime and the $Y$ component of
$F_{H4 \perp 4}(X,Y)$ is in $\perp 4$, then

\vspace{12pt}

the Lie bracket product $ [X,Y] = 0$
so that $\Lambda([X,Y]) = 0$
and therefore

$F_{H4 \perp 4}(X,Y) = (1/2) [\Lambda(X),\Lambda(Y)] =
(1/2) \partial_{X} \Lambda(Y) $.

\vspace{12pt}

The total value of $F_{H4 \perp 4}(X,Y)$ is then
$F_{H4 \perp 4}(X,Y) = \partial_{X}\Lambda(Y) $.

\vspace{12pt}

Integration of $\Lambda$ over $\perp 4$ gives

$\int _{Y \epsilon \perp 4}  \partial_{X}\Lambda(Y) =
\partial_{X}\Phi$,

where, as above, $\Phi$ is a 0-form taking values in the
$SU(2)$ Lie algebra.

\vspace{12pt}

As above, the $SU(2)$ values of $\Phi$ are represented by
complex
$SU(2)=Spin(3)$ doublets $\Phi = (\Phi^{+}, \Phi^{0})$.

\vspace{12pt}

In real terms, $\Phi^{+} = (\Phi_{1} + i \Phi_{2}) /
\sqrt{2}$ and
$\Phi^{0} = (\Phi_{3} + i \Phi_{4}) / \sqrt{2}$,

so $\Phi$ has 4 real degrees of freedom.

\vspace{12pt}

As discussed above, in the unitary gauge,
$ \Phi_{1} = \Phi_{2} = \Phi_{4} = 0$, and

$\Phi = (\Phi^{+}, \Phi^{0}) =
(1/ \sqrt{2})(\Phi_{1} + i \Phi_{2},
\Phi_{3} + i \Phi_{4}) = (1 / \sqrt{2})(0, v + H)$,

where $\Phi_{3} = (v + H) / \sqrt{2}$ ,

$v$ is the Higgs potential vacuum expectation value, and

$H$ is the real surviving Higgs scalar field.

\vspace{12pt}

The second term is then:

$\int  (\int_{\perp 4} - F_{H4 \perp 4} \wedge
\star F_{H4 \perp 4}) =$

$= \int (\int_{\perp 4} (-1/2) [\Lambda(X),\Lambda(Y)] \wedge
\star [\Lambda(X),\Lambda(Y)] ) = \int \partial
\overline{\Phi} \wedge \star \partial \Phi$

\vspace{12pt}

where the $SU(2)$ covariant derivative $\partial$ is

$\partial = \partial + \sqrt{\alpha_{w}} (W_{+} +
W_{-}) + \sqrt{\alpha_{w}} \cos{\theta_{w}}^{2} W_{0}$,
and $\theta_{w}$ is the Weinberg angle.

\vspace{12pt}

Then $\partial \Phi = \partial (v + H) /
 \sqrt{2} =$

$= [\partial H + \sqrt{\alpha_{w}} W_{+} (v + H) +
 \sqrt{\alpha_{w}} W_{-} (v + H) + \sqrt{\alpha_{w}} W_{0}
 (v + H) ] / \sqrt{2}$.

\vspace{12pt}

In the $D_{4}-D_{5}-E_{6}$ model the $W_{+}$, $W_{-}$,
 $W_{0}$, and $H$ terms are considered to be linearly
 independent.

\vspace{12pt}

$v = v_{+} + v_{-} + v_{0}$ has linearly
 independent components  $v_{+}$, $v_{-}$, and $v_{0}$ for
 $W_{+}$, $W_{-}$, and $W_{0}$.

\vspace{12pt}

$H$ is the Higgs component.

\vspace{12pt}

$\partial \overline{\Phi} \wedge \star \partial \Phi$ is
the sum of the squares of the individual terms.

\vspace{12pt}

Integration over $\perp 4$ involving two derivatives
$\partial_{X} \partial_{X}$ is taken to
change the sign by $i^{2} = -1$.

\vspace{12pt}

Then:

$\partial \overline{\Phi} \wedge \star \partial \Phi =
(1/2) (\partial H)^{2} +$

$+ (1/2) [ \alpha_{w} v_{+}^{2} \overline{W_{+}} W_{+} +
  \alpha_{w} v_{-}^{2} \overline{W_{-}} W_{-} +
  \alpha_{w} v_{0}^{2} \overline{W_{0}} W_{0} ] +$

$+ (1/2) [ \alpha_{w} \overline{W_{+}} W_{+} +
  \alpha_{w} \overline{W_{-}} W_{-} +
  \alpha_{w} \overline{W_{0}} W_{0} ] [ H^{2} + 2 v H ]$.

\vspace{12pt}

Then the full curvature term of the weak-Higgs Lagrangian,

$\int F_{w} \wedge \star F_{w} +  \partial
\overline{\Phi} \wedge \star \partial \Phi +
\lambda (\overline{\Phi} \Phi)^{2}  -
\mu^{2} \overline{\Phi} \Phi$,

\vspace{12pt}

is, by the Higgs mechanism:

\vspace{12pt}

$\int [ F_{w} \wedge \star F_{w}  +$

$+ (1/2) [ \alpha_{w} v_{+}^{2} \overline{W_{+}} W_{+} +
  \alpha_{w} v_{-}^{2} \overline{W_{-}} W_{-} +
  \alpha_{w} v_{0}^{2} \overline{W_{0}} W_{0} ] +$

$+ (1/2) [  \alpha_{w} \overline{W_{+}} W_{+} +
  \alpha_{w} \overline{W_{-}} W_{-} +
  \alpha_{w} \overline{W_{0}} W_{0} ] [ H^{2} + 2 v H ] +$

$+ (1/2) (\partial H)^{2}  + (1/4) \mu^{2} H^{2}  -$

$- (1/16) \mu^{2} v^{2} [ 1 - 4H^{3} / v^{3} - H^{4} /
 v^{4} ]  ] $.

\vspace{12pt}

The weak boson Higgs mechanism masses, in terms of
$v = v_{+} + v_{-} + v_{0}$, are:

\vspace{12pt}

$(\alpha_{w} / 2) v_{+}^{2} = m_{W_{+}}^{2}$ ;

\vspace{12pt}

$(\alpha_{w} / 2) v_{-}^{2} = m_{W_{-}}^{2}$ ;  and

\vspace{12pt}

$(\alpha_{w} / 2) v_{0}^{2} = m_{W_{+0}}^{2}$,

\vspace{12pt}

with $( v = v_{+} + v_{-} + v_{0} ) =  ((\sqrt{2}) /
\sqrt{\alpha_{w}}) ( m_{W_{+}} +  m_{W_{-}} + m_{W_{0}} )$.

\vspace{12pt}

\newpage

Then:

\vspace{12pt}

$\int [ F_{w} \wedge \star F_{w}  +$

$+ m_{W_{+}}^{2} W_{+} W_{+} +   m_{W_{-}}^{2} W_{-} W_{-} +
  m_{W_{0}}^{2} W_{0} W_{0}  +$

$+ (1/2) [ \alpha_{w} \overline{W_{+}} W_{+} +
  \alpha_{w} \overline{W_{-}} W_{-} +
  \alpha_{w} \overline{W_{0}} W_{0} ] [ H^{2} + 2vH ] +$

$+ (1/2)(\partial H)^{2} + (1/2)(\mu^{2} / 2)H^{2} -$

$- (1/16 \mu^{2} v^{2} [1 - 4H^{3} / v^{3} -
H^{4} / v^{4}]$.

\vspace{12pt}

\newpage

\subsection{Gauge Boson Part of the Lagrangian.}

In this subsection, we will look at matrix
representations of $Spin(0,8)$ and how dimensional
reduction affects them.

\vspace{12pt}

For a similar study from the point of view of
Clifford algebras, see
\newline
\href{http://xxx.lanl.gov/abs/hep-th/9402003}{hep-th/9402003}
\cite{SMI3} ).
\newline
For errata for that paper (and others), see
\newline
\href{http://www.gatech.edu/tsmith/Errata.html}{WWW URL
http://www.gatech.edu/tsmith/Errata.html}
\cite{SMI6} ).

\vspace{12pt}

\vspace{12pt}

The gauge boson bivector part of the Lagrangian is
\begin{equation}
\int_{V_{8}} F_{8} \wedge \star F_{8}
\end{equation}

It represents the $D_{4}-D_{5}-E_{6}$ model gauge
group $Spin(0,8)$ acting in 8-dimensional spacetime.
\newline
\vspace{12pt}

The $8 \times 8$ matrix representation of the $Spin(0,8)$
Lie algebra with the commutator bracket product [,] is

\begin{equation}
\left( \begin{array} {cccccccc}
      0 & a_{12} & a_{13} & a_{14}  & a_{15} & a_{16} & a_{17} & a_{18}\\
-a_{12} &      0 & a_{23} & a_{24}  & a_{25} & a_{26} & a_{27} & a_{28}\\
-a_{13} &-a_{23} &      0 & a_{34}  & a_{35} & a_{36} & a_{37} & a_{38}\\
-a_{14} &-a_{24} &-a_{34} &      0  & a_{45} & a_{46} & a_{47} & a_{48}\\
-a_{15} &-a_{25} &-a_{35} &-a_{45}  &      0 & a_{56} & a_{57} & a_{58}\\
-a_{16} &-a_{26} &-a_{36} &-a_{46}  &-a_{56} &      0 & a_{67} & a_{68}\\
-a_{17} &-a_{27} &-a_{37} &-a_{47}  &-a_{57} &-a_{67} &      0 & a_{78}\\
-a_{18} &-a_{28} &-a_{38} &-a_{48}  &-a_{58} &-a_{68} &-a_{78} &      0\\
\end{array} \right)
\end{equation}

To see how the $Spin(0,8)$ is affected by dimensional
reduction of spacetime to 4 dimensions, represent
$Spin(0,8)$, as in section 3.2, by

\begin{equation}
\left(
\begin{array}{cc}
S^{7}_{1} & 0\\
& \\
0 & S^{7}_{2}
\end{array}
\right)
\oplus G_{2}
\end{equation}
This Lie algebra is 7+7+14 = 28-dimensional $Spin(0,8)$,
also denoted $D_{4}$.
\newline
\vspace{12pt}

Recall that $S^{7}_{1}$ and $S^{7}_{2}$ represent
\newline
the imaginary octonions
$\{ e_{1},e_{2},e_{3},e_{4},e_{5},e_{6},e_{7} \}$.
\newline
{}From our present local Lie algebra point of view, they look
like linear tangent spaces to 7-spheres, not like
global round nonlinear 7-spheres.
\newline
Therefore, instead of using the Hopf fibration
$S^{3} \rightarrow S^{7} \rightarrow S^{4}$,
\newline
we break the spaces down in accord with dimensional reduction to:

\begin{equation}
S^{7}_{1} \rightarrow {\bf R}^{3}_{1} \oplus {\bf R}^{4}_{1}
\end{equation}

and

\begin{equation}
S^{7}_{2} \rightarrow {\bf R}^{3}_{2} \oplus {\bf R}^{4}_{2}
\end{equation}

\vspace{12pt}

$G_{2}$ has two fibrations:

\begin{equation}
SU(3) \rightarrow G^{2} \rightarrow S^{6}.
\end{equation}

\begin{equation}
SU(2) \otimes SU(2) \rightarrow G^{2} \rightarrow
M(G_{2})_{8}
\end{equation}
where $M(G_{2})_{8}$ is an 8-dimensional
homogeneous rank 2 symmetric space.
\newline
\vspace{12pt}

By choice of which $G_{2}$ fibration to use,
$Spin(0,8)$ has two decompositions from octonionic
derivations $G_{2}$ to quaternionic derivations $SU(2)$.
\newline
\vspace{12pt}

First, choose the $SU(3)$ subgroup of $G_{2}$ by
choosing the $G_{2}$ fibration
\begin{equation}
SU(3) \rightarrow G^{2} \rightarrow S^{6}.
\end{equation}
Since the $SU(3)$ subgroup of $G_{2}$ is the
larger 8-dimensional part of 14-dimensional $G_{2}$,
also take the larger ${\bf R}^{4}$ parts of the $S^{7}$'s, to get:

\begin{equation}
\left(
\begin{array}{cc}
{\bf R}^{4}_{1} & 0\\
& \\
0 & {\bf R}^{4}_{2}
\end{array}
\right)
\oplus SU(3)
\end{equation}
\newline
\vspace{12pt}

If ${\bf R}^{4}_{1} \oplus {\bf R}^{4}_{2}$ is identified with
\newline
the local tangent space of the 8-dimensional manifold
$$(Spin(0,6) \times U(1)) / SU(3)$$
then we have constructed
\newline
the $Spin(0,6) \times U(1)$ subgroup of $Spin(0,8)$.

\vspace{12pt}

The 8-dimensional manifold $(Spin(0,6) \times U(1)) / SU(3)$
\newline
is built up from the 6-dimensional irreducible symmetric space
$$Spin(0,6) / U(3)$$
(see the book $Einstein \; Manifolds$ by Besse \cite{BES})
\newline
 by adding two $U(1)$'s, one $U(1) = U(3) / SU(3)$ and
\newline
the other the $U(1)$ in $Spin(0,6) \times U(1)$.

\vspace{12pt}

Now that we have built $Spin(0,6) \times U(1)$ from
dimensional reduction process acting on $Spin(0,8)$,
compare an $8 \times 8$ matrix representation:

\begin{equation}
\left( \begin{array} {cccccccc}
      0 & a_{12} & a_{13} & a_{14}  & a_{15} & a_{16} & 0      & 0     \\
-a_{12} &      0 & a_{23} & a_{24}  & a_{25} & a_{26} & 0      & 0     \\
-a_{13} &-a_{23} &      0 & a_{34}  & a_{35} & a_{36} & 0      & 0     \\
-a_{14} &-a_{24} &-a_{34} &      0  & a_{45} & a_{46} & 0      & 0     \\
-a_{15} &-a_{25} &-a_{35} &-a_{45}  &      0 & a_{56} & 0      & 0     \\
-a_{16} &-a_{26} &-a_{36} &-a_{46}  &-a_{56} &      0 & 0      & 0     \\
 0      & 0      & 0      & 0       & 0      & 0      &      0 & a_{78}\\
 0      & 0      & 0      & 0       & 0      & 0      &-a_{78} &      0\\
\end{array} \right)
\end{equation}

\vspace{12pt}

\newpage

Now make the second choice, the $SU(2) \times SU(2)$
subgroup of $G_{2}$ by choosing the $G_{2}$ fibration

\begin{equation}
SU(2) \otimes SU(2) \rightarrow G^{2} \rightarrow
M(G_{2})_{8}
\end{equation}
where $M(G_{2})_{8}$ is an 8-dimensional
homogeneous rank 2 symmetric space.
\newline
\vspace{12pt}

Since the $SU(2) \times SU(2)$ subgroup of $G_{2}$ is the
smaller 6-dimensional part of 14-dimensional $G_{2}$,
also take the smaller ${\bf R}^{3}$ parts of the $S^{7}$'s,
to get:

\begin{equation}
\left(
\begin{array}{cc}
{\bf R}^{3}_{1} & 0\\
& \\
0 & {\bf R}^{3}_{2}
\end{array}
\right)
\oplus SU(2) \oplus SU(2)
\end{equation}
\vspace{12pt}

If ${\bf R}^{3}_{1} \oplus {\bf R}^{3}_{2}$ is identified with
\newline
the local tangent space of the 6-dimensional manifold
$$U(3) / Spin(3)$$
then, since $Spin(3) = SU(2)$,  we have constructed
\newline
a $U(3) \times SU(2)$ subgroup of $Spin(0,8)$.

\vspace{12pt}

The 6-dimensional manifold $U(3) / Spin(3)$
\newline
is built up from the 5-dimensional irreducible symmetric space
$$SU(3) / Spin(3)$$
(see the book $Einstein \; Manifolds$ by Besse \cite{BES})
\newline
 by adding the $U(1)$ from $U(1) = U(3) / SU(3)$.

\vspace{12pt}

Now that we have built $U(3) \times SU(2)$ from the
\newline
dimensional reduction process acting on $Spin(0,8)$,
\newline
note that $U(3)$ = $SU(3) \times U(1)$ so that
\newline
we have the 12-dimensional Standard Model gauge group
$$SU(3) \times SU(2) \times U(1)$$
Actually, it is even nicer to say that it is
$$U(3) \times SU(2)$$
by putting the electromagnetic $U(1)$ with the color
force $SU(3)$, because,
\newline
as O'Raifeartaigh says in section 9.4 of his book
\newline
$Group \; Structure \; of \; Gauge \; Theories$ \cite{ORA},
\newline
that is the most natural representation of
the Standard Model gauge groups.
\newline
This is because, when fermion representations
are taken into account,
\newline
the unbroken symmetry is
the $U(3)$ of electromagnetism and the color force,
\newline
while the weak force $SU(2)$ is broken by the Higgs mechanism.
\newline
\vspace{12pt}

One of the few differences between the Standard Model
sector of the 4-dimensional $D_{4}-D_{5}-E_{6}$ model
and the Standard Model electroweak structure is that
\newline
in the $D_{4}-D_{5}-E_{6}$ model the electromagnetic
$U(1)$ is most naturally put with the color force $SU(3)$,
while
\newline
in the electroweak Standard Model the electromagnetic $U(1)$
is most naturally put with the weak force $SU(2)$.
\newline
As O'Raifearteagh \cite{ORA} points out, that difference
is an advantage of the $D_{4}-D_{5}-E_{6}$ model.

\vspace{12pt}

Now, look at the Standard Model sector of $D_{4}-D_{5}-E_{6}$ model
from the matrix representation point of view:

\vspace{12pt}

Consider a $U(4)$ subalgebra of $Spin(0,8)$.

\vspace{12pt}

$U(4)$ can be represented (see section 412 G of \cite{EDM})
as a subalgebra of $Spin(0,8)$ by

\begin{equation}
\begin{array}{|c|c|}
\hline
{\bf{Re}}(U_{4}) &{\bf{Im}}(U_{4})  \\
\hline
{\bf{-Im}}(U_{4}) & {\bf{Re}}(U_{4})  \\
\hline
\end{array}
\end{equation}

\vspace{12pt}

Therefore, the $U(4)$ subalgebra of $Spin(0,8)$ can be represented by

\begin{equation}
\left( \begin{array} {cccccccc}
      0 & u_{12} & u_{13} & u_{14} & v_{11} & v_{12} & v_{13} & v_{14} \\
-u_{12} &      0 & u_{23} & u_{24} & v_{12} & v_{22} & v_{23} & v_{24} \\
-u_{13} &-u_{23} &      0 & u_{34} & v_{13} & v_{23} & v_{33} & v_{34} \\
-u_{14} &-u_{24} &-u_{34} &      0 & v_{14} & v_{24} & v_{34} & v_{44} \\
-v_{11} &-v_{12} &-v_{13} &-v_{14} &      0 & u_{12} & u_{13} & u_{14}\\
-v_{12} &-v_{22} &-v_{23} &-v_{24} &-u_{12} &      0 & u_{23} & u_{24}\\
-v_{13} &-v_{23} &-v_{33} &-v_{34} &-u_{13} &-u_{23} &      0 & u_{34}\\
-v_{14} &-v_{24} &-v_{34} &-v_{44} &-u_{14} &-u_{24} &-u_{34} &      0\\
\end{array} \right)
\end{equation}
\newline
\vspace{12pt}

(Compare this representation of $U(4)$ with the representation above
of the isomorphic algebra $Spin(0,6) \times U(1)$.)
\newline
\vspace{12pt}

A 12-dimensional subalgebra is

\begin{equation}
\left( \begin{array} {cccccccc}
      0 & u_{12} & u_{13} & u_{14} &      0 & v_{12} & v_{13} & v_{14} \\
-u_{12} &      0 & u_{23} & u_{24} & v_{12} &      0 & v_{23} & v_{24} \\
-u_{13} &-u_{23} &      0 & u_{34} & v_{13} & v_{23} &      0 & v_{34} \\
-u_{14} &-u_{24} &-u_{34} &      0 & v_{14} & v_{24} & v_{34} &      0 \\
      0 &-v_{12} &-v_{13} &-v_{14} &      0 & u_{12} & u_{13} & u_{14} \\
-v_{12} &      0 &-v_{23} &-v_{24} &-u_{12} &      0 & u_{23} & u_{24} \\
-v_{13} &-v_{23} &      0 &-v_{34} &-u_{13} &-u_{23} &      0 & u_{34} \\
-v_{14} &-v_{24} &-v_{34} &      0 &-u_{14} &-u_{24} &-u_{34} &      0 \\
\end{array} \right)
\end{equation}
\newline
\vspace{12pt}

\newpage

As we have seen in this subsection, the 28 infinitesimal
generators of $Spin(8)$ are broken by dimensional reduction
into two parts:
\newline
\vspace{12pt}

16-dimensional $U(4)$ = $Spin(0,6) \times U(1)$, where
$Spin(0,6)$ = $SU(4)$ is the conformal group of 4-dimensional
spacetime (the conformal group gives gravity by the
MacDowell-Mansouri mechanism \cite{MAC,MOH},
and gauge fixing of the
conformal group gives the Higgs scalar symmetry breaking)
and the $U(1)$ is the complex phase propagators
in the 4-dimensional spacetime; and
\newline
\vspace{12pt}

Physically, the $U(1)$ is the Dirac complexification and
it gives the physical Dirac gammas their complex structure,
so that they are ${\bf C}(4)$ instead of ${\bf R}(4)$.

\vspace{12pt}

Dirac complexification justifies the physical use of
Wick rotation between Euclidean and Minkowski spacetimes,
because ${\bf{C}}(4)$ is the Clifford algebra of both
the compact Euclidean deSitter Lie group $Spin(0,5)$
and the non-compact Minkowski anti-deSitter Lie group
$Spin(2,3)$.
\newline
\vspace{12pt}

12-dimensional $U(3) \times SU(2)$, where $SU(2)$ is the
gauge group of the weak force and $U(3)$ = $SU(3) \times U(1)$
is the $SU(3)$ gauge group of the color force and the
$U(1)$ gauge group of electromagnetism.

\newpage

\subsubsection{Conformal Gravity and Higgs Scalar.}
$Spin(0,6)$ is the maximal subgroup of $Spin(0,8)$ that
\newline
acts on the 4-dimensional reduced spacetime.
\newline
It acts as the compact version of the conformal group.
\newline
An $8 \times 8$ matrix representation of $Spin(0,6)$ is

\begin{equation}
\left( \begin{array} {cccccccc}
      0 & a_{12} & a_{13} & a_{14}  & a_{15} & a_{16} & 0      & 0     \\
-a_{12} &      0 & a_{23} & a_{24}  & a_{25} & a_{26} & 0      & 0     \\
-a_{13} &-a_{23} &      0 & a_{34}  & a_{35} & a_{36} & 0      & 0     \\
-a_{14} &-a_{24} &-a_{34} &      0  & a_{45} & a_{46} & 0      & 0     \\
-a_{15} &-a_{25} &-a_{35} &-a_{45}  &      0 & a_{56} & 0      & 0     \\
-a_{16} &-a_{26} &-a_{36} &-a_{46}  &-a_{56} &      0 & 0      & 0     \\
 0      & 0      & 0      & 0       & 0      & 0      &      0 &      0\\
 0      & 0      & 0      & 0       & 0      & 0      &      0 &      0\\
\end{array} \right)
\end{equation}

\vspace{12pt}

The MacDowell-Mansouri mechanism \cite{MAC} produces a classical model
of gravity from a $Spin(0,5)$ de Sitter gauge group.  As a subalgebra
of the $Spin(0,6)$ Lie algebra, the $Spin(0,5)$ Lie algebra can be
represented by

\begin{equation}
\left( \begin{array} {cccccccc}
      0 & a_{12} & a_{13} & a_{14}  & a_{15} &      0 & 0      & 0     \\
-a_{12} &      0 & a_{23} & a_{24}  & a_{25} &      0 & 0      & 0     \\
-a_{13} &-a_{23} &      0 & a_{34}  & a_{35} &      0 & 0      & 0     \\
-a_{14} &-a_{24} &-a_{34} &      0  & a_{45} &      0 & 0      & 0     \\
-a_{15} &-a_{25} &-a_{35} &-a_{45}  &      0 &      0 & 0      & 0     \\
      0 &      0 &      0 &      0  &      0 &      0 & 0      & 0     \\
 0      & 0      & 0      & 0       & 0      & 0      &      0 &      0\\
 0      & 0      & 0      & 0       & 0      & 0      &      0 &      0\\
\end{array} \right)
\end{equation}

\vspace{12pt}

The 5 elements $\{ a_{16}, a_{26},a_{36},a_{46},a_{56} \}$ that are
in $Spin(0,6)$ but not in $Spin(0,5)$ are the 1 scale and 4 conformal
degrees of freedom that are gauge-fixed by Mohapatra in
section 14.6 of \cite{MOH} to get from
the 15-dimensional conformal group $Spin(0,6)$ to the
10-dimensional deSitter group $Spin(0,5)$ so that the
MacDowell-Mansouri mechanism \cite{MAC} can be used to produce
Einstein-Hilbert gravity plus a cosmological constant term,
an Euler topological term, and a Pontrjagin topological term.
\newline
As Nieto, Obregon, and Socorro \cite{NIE} have shown,
MacDowell-Mansouri deSitter gravity is equivalent to
Ashtekar gravity plus a cosmological constant term,
an Euler topological term, and a Pontrjagin topological term.

\vspace{12pt}

For further discussion of the MacDowell-Mansouri mechanism,
\newline
see Freund \cite{FRE} (chapter 21),
or Ne'eman and Regge \cite{NEE}(at pages 25-28),
\newline
or \href{http://xxx.lanl.gov/abs/gr-qc/9402029}
{Nieto, Obregon, and Socorro} \cite{NIE}

\vspace{12pt}

 The physical reason for fixing the scale and conformal degrees of
freedom lies in the relationship between gravity and
the Higgs mechanism.

\vspace{12pt}

Since all rest mass comes from the Higgs mechanism,
and since rest mass interacts through gravity,
it is natural for gravity and Higgs
symmetry breaking to be related at a fundamental level.

\vspace{12pt}

As remarked by
\href{http://xxx.lanl.gov/abs/gr-qc/9410045}{Sardanashvily}
\cite{SAR4},
Heisenberg and Ivanenko in the 1960s made the first atttempt
to connect gravity
with a symmetry breaking mechanism by proposing that
the graviton might be
a Goldstone boson resulting from breaking Lorentz symmetry
in going from
flat MInkowski spacetime to curved spacetime.

\vspace{12pt}

\href{http://xxx.lanl.gov/abs/gr-qc/9410045}{Sardanashvily}
\cite{SAR4}
(see also
\href{http://xxx.lanl.gov/abs/gr-qc/9405013}{gr-qc/9405013}
\cite{SAR2}
\nolinebreak
\href{http://xxx.lanl.gov/abs/gr-qc/9407032}{gr-qc/9407032}
\cite{SAR3}
\newline
\href{http://xxx.lanl.gov/abs/gr-qc/9411013}{gr-qc/9411013}
\cite{SAR5})
proposes that gravity be represented by a gauge theory
with group $GL(4)$, that $GL(4)$ symmetry can be broken
to either Lorentz $SO(3,1)$ symmetry or $SO(4)$ symmetry,
and that the resulting Higgs fields can be interpreted
as either the gravitational field (for breaking to $SO(3,1)$ or
the Riemannian metric (for breaking to $SO(4)$.

\vspace{12pt}

The identification of a pseudo-Riemannian metric with
a Higgs field was made by Trautman \cite{TRA},
by Sardanashvily \cite{SAR1} , and
by Ivanenko and Sardanashvily \cite{IVA}

\vspace{12pt}

In the $D_{4}-D_{5}-E_{6}$ model (using here the compact version)
the conformal group $Spin(0,6)$ = $SU(4)$ is broken to
the de Sitter group $Spin(0,5)$ = $Sp(2)$ by fixing
the 1 scale and 4 conformal gauge degrees of freedom.

\vspace{12pt}

The resulting Higgs field is interpreted in the
$D_{4}-D_{5}-E_{6}$ model as the
same Higgs field that gives mass to the $SU(2)$
weak bosons and to the Dirac fermions by the Higgs mechanism.

\vspace{12pt}

The Higgs mechanism requires "spontaneous symmetry
breaking" of a scalar field potential whose minima
are not zero, but which form a 3-sphere $S^{3}$ = $SU(2)$.

\vspace{12pt}

In particular, one real component of the complex
Higgs scalar doublet is set to $v / \sqrt{2)}$,
where $v$ is the modulus of the $S^{3}$ of minima,
usually called the vacuum expectation value.

\vspace{12pt}

If the $S^{3}$ is taken to be the unit quaternions,
then the "spontaneous symmetry breaking"
requires choosing a (positive) real axis for the
quaternion space.

\vspace{12pt}

In the standard model, it is assumed that a random
vacuum fluctuation breaks the $SU(2)$ symmetry and
in effect chooses a real axis at random.

\vspace{12pt}

 In the $D_{4}-D_{5}-E_{6}$ model, the symmetry breaking
from conformal $Spin(0,6)$ to de Sitter $Spin(0,5)$ by
fixing the 1 scale and 4 conformal gauge
degrees of freedom is a symmetry breaking
mechanism that does not require perturbation by
a random vacuum fluctuation.

\vspace{12pt}

Gauge-fixing the 1 scale degree of freedom fixes a length scale.
It can be chosen to be the magnitude of the vacuum expectation value,
or radius of the $S^{3}$.

\vspace{12pt}

Gauge-fixing the 4 conformal degrees of freedom fixes the
(positive) real axis of the $S^{3}$ consistently
throughout 4-dimensional  spacetime.

\vspace{12pt}

Therefore, the $D_{4}-D_{5}-E_{6}$ model Higgs field comes from the
breaking of $Spin(0,6)$ conformal symmetry to $Spin(0,5)$ de Sitter
gauge symmetry, from which Einstein-Hilbert gravity can
be constructed by the MacDowell-Mansouri mechanism.

\vspace{12pt}

Einstein-Hilbert gravity as a spin-2 field theory in flat spacetime:
Feynman, in his 1962-63 lectures at Caltech \cite{FEY},
showed how Einstein-Hilbert gravity can be described by
starting with a linear spin-2 field theory in flat spacetime,
and then adding higher-order terms to get Einstein-Hilbert
gravity.  The observed curved spacetime is based on an
unobservable flat spacetime.
(see also Deser \cite{DES}

\vspace{12pt}

The Feynman spin-2 flat spacetime construction of Einstein-Hilbert
gravity allows the $D_{4}-D_{5}-E_{6}$ model to be based
on a fundamental $D_{4}$ lattice 4-dimensional spacetime.

\vspace{12pt}

Quantization in the $D_{4}-D_{5}-E_{6}$ model is
fundamentally based on a path integral sum over
histories of paths in a $D_{4}$ lattice spacetime
using a generalized Feynman checkerboard.

\vspace{12pt}

The generalized Feynman checkerboard is discussed at
\newline
\href{http://www.gatech.edu/tsmith/FynCkbd.html}
\newline
{WWW URL http://www.gatech.edu/tsmith/FynCkbd.html} \cite{SMI6}.

\vspace{12pt}

Fundamentally, that is nice, but calculations can
be very difficult, particularly for quantum gravity.
\newline
\vspace{12pt}

The work of \href{http://xxx.lanl.gov/abs/hep-th/9408003}
{Garcia-Compean et. al} \cite{GAR} suggests that the most practical
approach to quantum gravity may be through BRST symmetry.

\vspace{12pt}

Quantization breaks the gauge group invariance of
the $D_{4}-D_{5}-E_{6}$ model Lagrangian,
because the path integral must not overcount paths
by including more than one representative of
each gauge-equivalence class of paths.
\newline
The remaining quantum symmetry is the symmetry of BRST
cohomology classes.
\newline
Knowledge of the BRST symmetry tells you which ghosts
must be used in quantum calculations,
so the BRST cohomology can be taken to be
the basis for the quantum theory.

\vspace{12pt}

\href{http://xxx.lanl.gov/abs/hep-th/9408003}
{Garcia-Compean et. al} \cite{GAR}
discuss two current approaches to quantum gravity:

\vspace{12pt}

string theory, which abandons point particles even
at the classical level;
and

\vspace{12pt}

redefinition of classical general relativity in terms
of new variables, the
\newline
\href{http://vishnu.nirvana.phys.psu.edu/newvariables.ps}
{Ashtekar variables} \cite{ASH}
and trying to use the new variables
to construct a quantum theory of gravity.

\vspace{12pt}

\href{http://xxx.lanl.gov/abs/gr-qc/9402029}
{Nieto, Obregon, and Socorro} \cite{NIE} have shown
\newline
that the MacDowell-Mansouri $Spin(0,5) = Sp(2)$
de Sitter Lagrangian for
\newline
gravity used in the $D_{4}-D_{5}-E_{6}$ model is equal to
\newline
the Lagrangian for gravity in terms of the
\newline
\href{http://vishnu.nirvana.phys.psu.edu/newvariables.ps}
{Ashtekar variables} \cite{ASH} plus

\vspace{12pt}

a cosmological constant term,
\newline
\vspace{12pt}

an Euler topological term, and

\vspace{12pt}

a Pontrjagin topological term.

\vspace{12pt}

Therefore, although the quantum gravity methods of string theory
cannot be used in the $D_{4}-D_{5}-E_{6}$ model
because the $D_{4}-D_{5}-E_{6}$ model
uses fundamental point particles at the classical level,

\vspace{12pt}

the methods based on
\href{http://vishnu.nirvana.phys.psu.edu/newvariables.ps}
{Ashtekar variables} \cite{ASH}
are available.

\vspace{12pt}

Two such approaches are:

\vspace{12pt}

a topological approach based on loop groups; and

\vspace{12pt}

an algebraic approach based on getting BRST transformations
\newline
from Maurer-Cartan horizontality conditions.

\vspace{12pt}

The latter approach, which is described in
\href{http://xxx.lanl.gov/abs/hep-th/9409046}{Blaga, et. al.}
\cite{BLA}
\newline
is the approach used for the $D_{4}-D_{5}-E_{6}$ model.

\vspace{12pt}

\subsubsection{Chern-Simons Time}
\vspace{12pt}

An essential part of a quantum theory of gravity
is the correct definition of physical time.

\vspace{12pt}

\href{http://xxx.lanl.gov/abs/gr-qc/9402029}
{Nieto, Obregon, and Socorro} \cite{NIE}
have shown that Lagrangian action of the
\href{http://vishnu.nirvana.phys.psu.edu/newvariables.ps}
{Ashtekar variables} \cite{ASH}
is a Chern-Simons action if the Killing metric of the
de Sitter group is used instead of the Levi-Civita tensor.

\vspace{12pt}

\href{http://xxx.lanl.gov/abs/gr-qc/9405015}{Smolin and Soo} \cite{SMO}
have shown that the Chern-Simons invariant of
the Ashtekar-Sen connection is a natural candidate
for the internal time coordinate
for classical and quantum cosmology,
so that the $D_{4}-D_{5}-E_{6}$ model
uses Chern-Simons time.

\vspace{12pt}

\subsubsection{Quantum Gravity plus Standard Model}

Another essential part of a quantum theory of gravity
is the correct relationship of quantum gravity with
the quantum theory of the forces and particles of the
Standard Model, to calculate how standard model particles
and fields interact in the presence of gravity.

\vspace{12pt}

\href{http://xxx.lanl.gov/abs/hep-th/9409081}{Moritsch, et. al.}
\cite{MOR} have done this by using Maurer-Cartan
horizontality conditions to get BRST transformations
for Yang-Mills gauge fields in the presence of gravity.

\vspace{12pt}

\newpage

\subsubsection{Color, Weak, and Electromagnetic Forces.}

12-dimensional $U(3) \times SU(2)$, where $SU(2)$ is the
gauge group of the weak force and $U(3)$ = $SU(3) \times U(1)$
is the $SU(3)$ gauge group of the color force and the
$U(1)$ gauge group of electromagnetism, can be represented
in terms of $8 \times 8$ matrices by

\begin{equation}
\left( \begin{array} {cccccccc}
      0 & u_{12} & u_{13} & u_{14} &      0 & v_{12} & v_{13} & v_{14} \\
-u_{12} &      0 & u_{23} & u_{24} & v_{12} &      0 & v_{23} & v_{24} \\
-u_{13} &-u_{23} &      0 & u_{34} & v_{13} & v_{23} &      0 & v_{34} \\
-u_{14} &-u_{24} &-u_{34} &      0 & v_{14} & v_{24} & v_{34} &      0 \\
      0 &-v_{12} &-v_{13} &-v_{14} &      0 & u_{12} & u_{13} & u_{14} \\
-v_{12} &      0 &-v_{23} &-v_{24} &-u_{12} &      0 & u_{23} & u_{24} \\
-v_{13} &-v_{23} &      0 &-v_{34} &-u_{13} &-u_{23} &      0 & u_{34} \\
-v_{14} &-v_{24} &-v_{34} &      0 &-u_{14} &-u_{24} &-u_{34} &      0 \\
\end{array} \right)
\end{equation}

\vspace{12pt}

In terms of the fibrations of $S^{7}$ and $G_{2}$, the 12-dimensional
subalgebra $U(3) \times SU(2)$ is represented by

\begin{equation}
\left(
\begin{array}{cc}
S^{3}_{a} & 0\\
& \\
0 & S^{3}_{b}
\end{array}
\right)
\oplus SU(2) \oplus SU(2)
\end{equation}

\vspace{12pt}

The result of this decomposition is that the gauge
boson bivector part of the 8-dimensional
$D_{4}-D_{5}-E_{6}$ Lagrangian
\begin{equation}
\int_{V_{8}} F_{Spin(0,8)} \wedge \star F_{Spin(0,8)}
\end{equation}
breaks down into
\begin{equation}
\int_{V_{4}} F_{Spin(0,6)} \wedge \star F_{Spin(0,6)} \oplus
F_{U(1)} \wedge \star F_{U(1)} \oplus
F_{SU(3)} \wedge \star F_{SU(3)} \oplus
F_{SU(2)} \wedge \star F_{SU(2)}
\end{equation}
The 28 gauge bosons of 8-dimensional $Spin(0,8)$ are broken
into four independent (or commuting, from the Lie algebra
point of view) sets of gauge bosons:

\vspace{12pt}

15 for gravity and Higgs symmetry breaking $Spin(0,6)$,
\newline
plus 1 for the $U(1)$ propagator phase;

\vspace{12pt}

1 for $U(1)$ electromagentism;

\vspace{12pt}

8 for color $SU(3)$; and

\vspace{12pt}

3 for the $SU(2)$ weak force.

\vspace{12pt}

Each of the terms of the form
\begin{equation}
\int_{V_{4}} F \wedge \star F
\end{equation}
contains a force strength constant.

\vspace{12pt}

The force strength constants define the
relative strengths of the four forces.

\vspace{12pt}

One of the factors determining the force strength
constants is the relative magnitude of the
measures of integration over the 4-dimensional
spacetime base manifold in each integral.

\vspace{12pt}

The relative magnitude of the measures is
proportional to the volume $Vol(M_{force})$ of the irreducible
$m_{force}$(real)-dimensional symmetric space on which
the gauge group acts naturally as a component of
4(real) dimensional spacetime
$M_{force}^{\left( 4 \over m_{force} \right)}$.

\vspace{12pt}

The $M_{force}$ manifolds for the gauge groups of
the four forces are:

\begin{equation}
\begin{array}{|c|c|c|c|}
\hline
Gauge \: Group & Symmetric \: Space & m_{force}
& M_{force}  \\
\hline
& & & \\
Spin(5) & Spin(5) \over Spin(4)  & 4 & S^4\\
& & & \\
SU(3) & SU(3) \over {SU(2) \times U(1)}
& 4  & {\bf C}P^2 \\
& & & \\
SU(2) & SU(2) \over U(1)  & 2 & S^2 \times S^2 \\
& & & \\
U(1) & U(1)  & 1 & S^1 \times S^1 \times S^1
\times S^1 \\
& & & \\
\hline
\end{array}
\end{equation}

Further discussion of this factor is in
\newline
\href{http://www.gatech.edu/tsmith/See.html}
\newline
{WWW URL http://www.gatech.edu/tsmith/See.html} \cite{SMI6}.

\vspace{12pt}

The second factor in the force strengths is based
on the interaction of the gauge bosons  with the
spinor fermions through the covariant derivative.

\vspace{12pt}

When the spinor fermion term is added to the
4-dimensional $D_{4}-D_{5}-E_{6}$ Lagrangian for each force,
you get a Lagrangian of the form
\begin{equation}
\int_{V_{4}} F \wedge \star F +
\overline{S_{8\pm}} \not \! \partial_{8} S_{8\pm}
\end{equation}
The covariant derivative part of the Dirac operator
$\not \! \partial_{8}$ gives the interaction between
each of the four forces and the spinor fermions.

\vspace{12pt}

The strength of each force depends on the magnitude of
the interaction of the covariant derivative of the force
with the spinor fermions.  Since the spinor fermions
are defined with respect to a
space $Q$ = $S^{7} \times {\bf R}P^{1}$ that is the Shilov
boundary of a bounded complex homogeneous domain $D$,
the relative strength of each force can be measured
by the relative volumes of the part of the manifolds
$Q$ and $D$ that are affected by that force.

\vspace{12pt}

Let $Vol(Q_{force})$ be the volume of that part of the full compact
fermion state space manifold ${\bf R}P^1 \times S^7$
on which a gauge group acts naturally through its
charged (color or electromagnetic charge) gauge bosons.

\vspace{12pt}

For the forces with charged gauge bosons,

\vspace{12pt}

$Spin(5)$ gravity,

\vspace{12pt}

$SU(3)$ color
force, and

\vspace{12pt}

$SU(2)$ weak force,

\vspace{12pt}

$Q_{force}$ is the Shilov boundary of
the bounded complex homogeneous
domain $D_{force}$ that corresponds to
the Hermitian symmetric
space on which the gauge group
acts naturally as a local isotropy (gauge) group.

\vspace{12pt}

For $U(1)$ electromagnetism, whose photon carries
no charge, the factors $Vol(Q_{U(1)})$ and $Vol(D_{U(1)})$ do
not apply and are set equal to $1$.

\vspace{12pt}

The volumes $Vol(M_{force})$, $Vol(Q_{force})$,
and $Vol(D_{force})$ are
calculated with $M_{force}, Q_{force}, D_{force}$
normalized to unit radius.

\vspace{12pt}

The factor
$1 \over {{Vol(D_{force})}^{\left( 1 \over m_{force} \right)}}$
is a normalization factor to be used if the dimension of
$Q_{force}$ is different from the dimension $m_{force}$,
in order to normalize the radius of $Q_{force}$ to
be consistent with the unit radius of $M_{force}$.

\vspace{12pt}

The $Q_{force}$ and $D_{force}$ manifolds for the gauge groups of
the four forces are:

\begin{equation}
\begin{array}{|c|c|c|c|c|}
\hline
Gauge & Hermitian & Type & m_{force}
& Q_{force}  \\
Group & Symmetric & of & & \\
& Space & D_{force} & & \\
\hline
& & & & \\
Spin(5) & Spin(7) \over {Spin(5) \times U(1)}
& IV_{5} &4 & {\bf R}P^1 \times S^4 \\
& & & & \\
SU(3) & SU(4) \over {SU(3) \times U(1)}
& B^6 \: (ball) &4 & S^5 \\
& & & & \\
SU(2) & Spin(5) \over {SU(2) \times U(1)}
& IV_{3} & 2 & {\bf R}P^1 \times S^2 \\
& & & & \\
U(1) & -  & - & 1  & - \\
& & & & \\
\hline
\end{array}
\end{equation}

\vspace{12pt}

The third factor affects only the force of gravity,
which has a characteristic mass because the Planck
length is the fundamental lattice length in the
$D_{4}-D_{5}-E_{6}$ model, so that $\mu_{Spin(0,5)} =
M_{Planck}$ and
\newline
the weak force, whose gauge bosons acquire mass
by the Higgs mechanism, so that $\mu_{Spin(0,5)} =
\sqrt{m_{W+}^{2} + m_{W-}^{2} + m_{W_{0}}^{2}}$.

\vspace{12pt}

For the weak force, the relevant factor is
$${1 \over \mu_{force}^2} = {1 \over {m_{W+}^{2}
+m_{W-}^{2} + m_{W_{0}}^{2}}}$$

\vspace{12pt}

For gravity, it is
$${1 \over \mu_{force}^2} = {1 \over {m_{Planck}^{2}}}$$

\vspace{12pt}

For the $SU(3)$ color and $U(1)$ electromagnetic forces,
$${1 \over \mu_{force}^2} = 1 $$

\vspace{12pt}

Taking all the factors into account, the calculated
strength of a force is taken to be
proportional to the product:
\begin{equation}
\left(1 \over \mu_{force}^2 \right) \left( Vol(M_{force})
\right)
\left( {Vol(Q_{force})} \over {{Vol(D_{force})}^{ \left( 1
\over m_{force} \right) }} \right)
\end{equation}

\vspace{12pt}

The geometric force strengths, that is,
\newline
everything but the mass scale factors $1 / \mu_{force}^{2}$,
\newline
normalized by dividing them by the largest one,
\newline
the one for gravity.

\vspace{12pt}

The geometric volumes needed for the calculations,
mostly taken from Hua \cite{HUA}, are

\begin{equation}
\begin{array}{||c||c|c||c|c||c|c||}
\hline
Force & M & Vol(M) & Q
& Vol(Q) & D & Vol(D  \\
\hline
& & & & & & \\
gravity & S^4 & 8\pi^{2}/3
& {\bf R}P^1 \times S^4  & 8\pi^{3}/3
& IV_{5} & \pi^{5}/2^{4} 5! \\
\hline
& & & & & &\\
color & {\bf C}P^2 & 8\pi^{2}/3
& S^5 & 4\pi^{3}
& B^6 \: (ball) & \pi^{3}/6 \\
\hline
& & & & & & \\
weak & {S^2} \times {S^2} & 2 \times {4 \pi}
& {\bf R}P^1 \times S^2 & 4 \pi^2
& IV_{3} & \pi^{3} / 24 \\
\hline
& & & & & & \\
e-mag  & T^4  & 4 \times {2\pi}
& -  & -
& -  & - \\
\hline
\end{array}
\end{equation}

Using these numbers, the results of the
calculations are the relative force strengths
at the characteristic energy level of the
generalized Bohr radius of each force:

\begin{equation}
\begin{array}{|c|c|c|c|c|}
\hline
Gauge \: Group & Force & Characteristic
& Geometric & Total \\
& & Energy & Force & Force \\
& & & Strength & Strength \\
\hline
& & & & \\
Spin(5) & gravity & \approx 10^{19} GeV
& 1 & G_{G}m_{proton}^{2} \\
& & & & \approx 5 \times 10^{-39} \\
\hline
& & & & \\
SU(3) & color & \approx 245 MeV & 0.6286
& 0.6286 \\
\hline
& & & & \\
SU(2) & weak & \approx 100 GeV & 0.2535
& G_{W}m_{proton}^{2} \approx  \\
& & & & \approx 1.02 \times 10^{-5} \\
\hline
& & & & \\
U(1) & e-mag  & \approx 4 KeV
& 1/137.03608  & 1/137.03608 \\
\hline
\end{array}
\end{equation}

The force strengths are given at the characteristic
energy levels of their forces, because the force
strengths run with changing energy levels.

\vspace{12pt}

The effect is particularly pronounced with the color
force.

\vspace{12pt}

In \href{http://www.gatech.edu/tsmith/cweRen.html}{WWW
URL http://www.gatech.edu/tsmith/cweRen.html} \cite{SMI6},
\newline
the color force strength was calculated
at various energies according to renormalization group
equations, with the following results:

\begin{equation}
\begin{array}{|c|c|}
\hline
Energy \: Level & Color \: Force \: Strength \\
\hline
&  \\
245 MeV & 0.6286 \\
& \\
5.3 GeV & 0.166 \\
& \\
34 GeV & 0.121  \\
& \\
91 GeV & 0.106 \\
& \\
\hline
\end{array}
\end{equation}
\newline
Shifman \href{http://xxx.lan.gov/abs/hep-ph/9501222}{WWW
URL http://xxx.lan.gov/abs/hep-ph/9501222} \cite{SHI}
has noted that Standard Model global fits at the $Z$ peak,
about $91 \; GeV$, give a color force strength of about
0.125 with $\Lambda_{QCD} \approx 500 \; MeV$,
\newline
whereas low energy results and lattice calculations
give  a color force strength at the $Z$ peak of about
0.11 with $\Lambda_{QCD} \approx 200 \; MeV$.

\vspace{12pt}

The low energy results and lattice calculations are
closer to the tree level $D_{4}-D_{5}-E_{6}$ model
value at $91 \; GeV$ of 0.106.
\newline
Also, the $D_{4}-D_{5}-E_{6}$ model has
$\Lambda_{QCD} \approx 245 \; MeV$
\newline
(For the pion mass, upon which the $\Lambda_{QCD}$
calculation depends, see
\newline
 \href{http://www.gatech.edu/tsmith/SnGdnPion.html}{WWW
URL http://www.gatech.edu/tsmith/SnGdnPion.html} \cite{SMI6}.)

\newpage

\subsection{Fermion Part of the Lagrangian}

Consider the spinor fermion term $\int {\overline{S_{8\pm}}
\not \!  \partial_{8} S_{8\pm}}$

\vspace{12pt}

For each of the surviving 4-dimensional $4$ and reduced
4-dimensional $\perp 4$ of 8-dimensional spacetime,
the part of $S_{8\pm}$ on which the Higgs $SU(2)$ acts
locally is $Q_{3} = {\bf{R}}P^{1} \times S^{2}$.

\vspace{12pt}

It is the Shilov boundary of the bounded domain $D_{3}$
that is isomorphic to the symmetric space
$\overline{D_{3}} = Spin(5)/SU(2) \times U(1)$.

\vspace{12pt}

The Dirac operator
$\not \!  \partial_{8} $ decomposes as
$\not \!  \partial = \not \!  \partial_{4}  +
\not \!  \partial_{\perp 4}$,
where

$\not \!  \partial_{4}$ is the Dirac operator
corresponding to the surviving spacetime $4$ and

$\not \!  \partial_{\perp 4}$ is the Dirac operator
corresponding to the reduced 4 $\perp 4$.

\vspace{12pt}

Then the spinor term is
  $\int {\overline{S_{8\pm}} \not \!  \partial_{4} S_{8\pm}} +
\overline{S_{8\pm}} \not \!  \partial_{\perp 4} S_{8\pm}$

The Dirac operator term   $\not \!  \partial_{\perp 4}$
in the reduced $\perp 4$ has dimension of mass.

\vspace{12pt}

After integration
$\int {\overline{S_{8\pm}} \not \!
\partial_{\perp 4} S_{8\pm}}$
over the reduced $\perp 4$,

$\not \! \partial_{\perp 4}$ becomes the real scalar
Higgs scalar field $Y = (v + H)$ that comes

from the complex $SU(2)$ doublet $\Phi$ after
action of the Higgs mechanism.

\vspace{12pt}

If integration over the reduced $\perp 4$ involving
two fermion terms $\overline{S_{8\pm}}$ and $S_{8\pm}$
is taken to change the sign by $i^{2} = -1$, then,
by the Higgs mechanism,

$\int \overline{S_{8\pm}} \not \! \partial_{\perp 4} S_{8\pm}
\rightarrow  \int(\int_{\perp 4}   \overline{S_{8\pm}}
\not \!  \partial_{\perp 4} S_{8\pm} )  \rightarrow  $

$\rightarrow - \int   \overline{S_{8\pm }} YY S_{8\pm }  =
- \int   \overline{S_{8\pm }} Y(v + H)  S_{8\pm }$,

\vspace{12pt}

where:

\vspace{12pt}

$H$ is the real physical Higgs scalar,
$m_{H} = v \sqrt(\lambda / 2)$, and
$v$ is the vacuum expectation value of the scalar field $Y$,
the free parameter in the theory that sets the mass scale.

\vspace{12pt}

Denote the sum of the three weak boson masses by
$\Sigma_{m_{W}}$.

\vspace{12pt}

$v = \Sigma_{m_{W}}((\sqrt{2}) / \sqrt{\alpha_{w}}) =
260.774 \times \sqrt{2} / 0.5034458 = 732.53 \; GeV$,

a value chosen so that the electron mass will be 0.5110 MeV.

\vspace{12pt}

The Higgs vacuum expectation value
$v = ( v_{+} + v_{-} + v_{0} )$
is the only particle mass free parameter.

\vspace{12pt}

In the $D_{4}-D_{5}-E_{6}$ model,
$v$ is set so that the electron mass $m_{e} = 0.5110 MeV$.

\vspace{12pt}

In the $D_{4}-D_{5}-E_{6}$ model, $\alpha_{w}$ is calculated
to be  $\alpha_{w} = 0.2534577$,

so $\sqrt{\alpha_{w}}$ = 0.5034458 and $v$ = 732.53 GeV.

\vspace{12pt}

The Higgs mass $m_{H}$ is given by the term
\begin{equation}
(1/2)(\partial H)^{2} - (1/2)(\mu^{2} / 2)H^{2} =
(1/2) [ (\partial H)^{2} -  (\mu^{2}/2) H^{2} ]
\end{equation}
to be
\begin{equation}
m_{H}^{2}  =  \mu^{2} / 2  =  \lambda v^{2} / 2
\end{equation}
so that
\begin{equation}
m_{H} = \sqrt{(\mu^{2} / 2)} =
\sqrt{\lambda} v^{2} / 2)
\end{equation}

\vspace{12pt}

$\lambda$ is the scalar self-interaction strength.
\newline
It should be the product of the "weak charges" of two
scalars coming from the reduced 4 dimensions in $Spin(4)$,
which should be the same as the weak charge of the surviving
weak force $SU(2)$ and therefore just the square of the
$SU(2)$ weak charge, $\sqrt{(\alpha_{w}^{2})} = \alpha_{w}$,
where $\alpha_{w}$ is the $SU(2)$ geometric force strength.

\vspace{12pt}

Therefore $\lambda = \alpha_{w}  = 0.2534576$,
$\sqrt{\lambda} = 0.5034458$, and $v$ = 732.53 GeV,

so that the mass of the Higgs scalar is
\begin{equation}
m_{H} = v \sqrt(\lambda / 2)  = 260.774 \; GeV.
\end{equation}

\vspace{12pt}

$Y$ is the Yukawa coupling between fermions and
the Higgs field.

\vspace{12pt}

$Y$ acts on all 28 elements (2 helicity states for each of
the 7 Dirac particles and 7 Dirac antiparticles)
of the Dirac fermions in a given generation,
because all of them are in the same $Spin(0,8)$
spinor representation.

\vspace{12pt}

\subsubsection{Calculation of Particle Masses.}

Denote the sum of the first generation Dirac fermion masses
by $\Sigma_{f_{1}}$.

\vspace{12pt}

Then $Y = (\sqrt{2}) \Sigma_{f_{1}} / v$, just as
$\sqrt(\alpha_{w}) = (\sqrt{2}) \Sigma_{m_{W}} / v$.

\vspace{12pt}

$Y$ should be the product of two factors:

\vspace{12pt}

$e^{2}$, the square of the electromagnetic charge
$e = \sqrt{\alpha_{E}}$ , because in the term
$\int(\int_{\perp 4} \overline{S_{8\pm }} \not \!
\partial _{\perp 4} S_{8\pm } )  \rightarrow
 - \int \overline{S_{8\pm }} Y(v + H)  S_{8\pm }$
each of the Dirac fermions $S_{8\pm}$ carries
electromagnetic charge proportional to $e$ ; and

\vspace{12pt}

$1/g_{w}$, the reciprocal of the weak
charge $g_{w} = \sqrt{\alpha_{w}}$,
because an $SU(2)$ force, the Higgs $SU(2)$, couples
the scalar field to the fermions.

\vspace{12pt}

Therefore
\begin{equation}
\Sigma_{f_{1}} = Y v / \sqrt{2} =
 (e^{2} / g_{w}) v / \sqrt{2}   = 7.508 \; GeV
\end{equation}
and

\begin{equation}
\Sigma_{f_{1}} / \Sigma_{m_{W}} = (e^{2} / g_{w}) v /
g_{w} v = e^{2} / g_{w}^{2} = \alpha_{E} / \alpha_{w}
\end{equation}

\vspace{12pt}

The Higgs term $- \int \overline{S_{8\pm}} Y(v + H)$
$S_{8\pm} = - \int \overline{S_{8\pm}}$
$Yv  S_{8\pm} - \int \overline{S_{8\pm}} YH S_{8\pm } = $
\
$= - \int \overline{S_{8\pm}} (\sqrt{2} \Sigma_{f_{1}})
S_{8\pm }  - \int \overline{S_{8\pm}} (\sqrt{2}
\Sigma_{f_{1}} / v)  S_{8\pm}$.

\vspace{12pt}

The resulting spinor term is of the form
$\int  [ \overline{S_{8\pm}} (\not \!  \partial - Yv)
S_{8\pm}    -  \overline{S_{8\pm}} YH S_{8\pm}  ]$
\newline
where $(\not \!  \partial - Yv)$ is a
massive Dirac operator.

\vspace{12pt}

How much of the total mass  $\Sigma_{f_{1}} =
Y v / \sqrt{2} = 7.5 \; GeV$ is allocated to each of the first
generation Dirac fermions is determined by calculating
the individual fermion masses in the $D_{4}-D_{5}-E_{6}$
model, and

\vspace{12pt}

those calculations also give the values of
\begin{equation}
\Sigma_{f_{2}} = 32.9 \; GeV
\end{equation}
\begin{equation}
\Sigma_{f_{3}} = 1,629 \; GeV
\end{equation}

as well as second and third generation individual
fermion masses, with the result that the individual
tree-level lepton masses and quark constituent masses are:

$m_{e}$ = 0.5110 MeV (assumed);
\newline
$m_{\nu_{e}}$ = $m_{\nu_{\mu}}$ = $m_{\nu_{\tau}}$ = 0;
\newline
$m_{d}$ = $m_{u}$ = 312.8 MeV  (constituent quark mass);
\newline
$m_{\mu}$ = 104.8 MeV;
\newline
$m_{s}$ =  625 MeV  (constituent quark mass);
\newline
$m_{c}$ =  2.09 GeV  (constituent quark mass);
\newline
$m_{\tau}$ =  1.88 GeV;
\newline
$m_{b}$ =  5.63 GeV  (constituent quark mass);
v
\begin{equation}
m_{t} \; = \; 130 \; GeV  \; (constituent \; quark \; mass).
\end{equation}

\vspace{12pt}

Here is how the individual fermion mass calculations
are done in the $D_{4}-D_{5}-E_{6}$ model.

\vspace{12pt}

The Weyl fermion neutrino has at tree level
only the left-handed state,
whereas the Dirac fermion electron and quarks can have
both left-handed and right-handed states,
so that the total number of states corresponding
to each of the half-spinor $Spin(0,8)$ representations is 15.

\vspace{12pt}

Neutrinos are massless at tree level in all generations.

\vspace{12pt}

In the  $D_{4}-D_{5}-E_{6}$model, the first generation
fermions correspond to octonions ${\bf O}$, while second
generation fermions correspond to pairs of octonions
${\bf O} \times {\bf O}$ and third generation fermions
correspond to triples of octonions
${\bf O} \times {\bf O} \times {\bf O}$.

\vspace{12pt}

To calculate the fermion masses in the model,
the volume of a compact manifold representing the
spinor fermions $S_{8+}$ is used.
It is the parallelizable manifold $S^7\times RP^1$.

\vspace{12pt}

 Also, since gravitation is coupled to mass,
the infinitesimal generators of the MacDowell-Mansouri
gravitation group, $Spin(0,5)$, are relevant.

\vspace{12pt}

The calculated quark masses are constituent masses, not
current masses.

\vspace{12pt}

In the $D_{4}-D_{5}-E_{6}$ model, fermion masses are
calculated as a product of four factors:
\begin{equation}
V(Q_{fermion}) \times N(Graviton) \times N(octonion) \times Sym
\end{equation}

$V(Q_{fermion})$ is the volume of the part of the half-spinor
\newline
fermion particle manifold $S^7\times RP^1$ that is
\newline
related to the fermion particle by photon,
weak boson, and gluon interactions.

\vspace{12pt}

$N(Graviton)$ is the number of types of $Spin(0,5)$ graviton related to
the fermion.  The 10 gravitons correspond to the 10 infinitesimal
generators of $Spin(0,5)$ = $Sp(2)$.
\newline
2 of them are in the Cartan subalgebra.
\newline
6 of them carry color charge, and may therefore be considered as
corresponding to quarks.
\newline
The remaining 2 carry no color charge, but
may carry electric charge and so may be considered as corresponding
to electrons.
\newline
One graviton takes the electron into itself, and the other can
only take the first-generation electron into the massless electron
neutrino.
\newline
Therefore only one graviton should correspond to the mass
of the first-generation electron.
\newline
The graviton number ratio of the down quark to the
first-generation electron is therefore 6/1 = 6.

\vspace{12pt}

$N(octonion)$ is an octonion number factor relating up-type quark
masses to down-type quark masses in each generation.

\vspace{12pt}

$Sym$ is an internal symmetry factor, relating  2nd and 3rd
generation massive leptons to first generation fermions.
\newline
It is not used in first-generation calculations.

\vspace{12pt}

The ratio of the down quark constituent mass to the electron mass
is then calculated as follows:
\newline
Consider the electron, e.
\newline
By photon, weak boson, and gluon interactions,
e can only be taken into 1, the massless neutrino.
\newline
The electron and neutrino, or their antiparticles,
cannot be combined to produce any of the
massive up or down quarks.
\newline
The neutrino, being massless at tree level,
does not add anything to the mass formula for the electron.
\newline
Since the electron cannot be related to any other massive Dirac
fermion, its volume $V(Q_{electron})$ is taken to be 1.

\vspace{12pt}

Next consider a red down quark $e_{3}$.
\newline
By gluon interactions, $e_{3}$ can be taken into
$e_{5}$ and $e_{7}$,
the blue and green down quarks.
\newline
By weak boson interactions, it can be taken into
$e_{1}$, $e_{2}$, and $e_{6}$, the red, blue, and green up quarks.
\newline
Given the up and down quarks, pions can be formed
from quark-antiquark pairs, and the pions can decay to produce
electrons and neutrinos.
\newline
Therefore the red down quark (similarly, any down quark)
is related to any part of $S^7\times {\bf R}P^1$,
the compact manifold corresponding to

$$\{ 1, e_{1}, e_{2}, e_{3}, e_{4}, e_{5}, e_{6}, e_{7} \}$$

and therefore a down quark should have a spinor manifold
volume factor $V(Q_{down quark}$ of the volume of
$S^7\times {\bf R}P^1$.
\newline
The ratio of the down quark spinor manifold volume factor to
the electron spinor manifold volume factor is just

\begin{equation}
V(Q_{down quark}) / V(Q_{electron}) =
V(S^7\times {\bf R}P^1)/1 = \pi ^{5} / 3.
\end{equation}

Since the first generation graviton factor is 6,

\begin{equation}
md/me = 6V(S^7 \times {\bf R}P^1) = 2 {\pi}^5 = 612.03937
\end{equation}

As the up quarks correspond to $e_{1}$, $e_{2}$, and $e_{6}$,
which are isomorphic to $e_{3}$, $e_{5}$, and $e_{7}$ of
the down quarks, the up quarks and down quarks
have the same constituent mass $m_{u} = m_{d}$.

\vspace{12pt}

Antiparticles have the same mass as the corresponding
particles.

\vspace{12pt}

Since the model only gives ratios of massses,
the mass scale is
fixed by assuming that the electron mass $m_{e}$ = 0.5110 MeV.

\vspace{12pt}

Then, the constituent mass of the down quark is
$m_{d}$ = 312.75 MeV, and
\newline
the constituent mass for the up quark is
$m_{u}$ = 312.75 MeV.

\vspace{12pt}

As the proton mass is taken to be the sum of the constituent
masses of its constituent quarks
\begin{equation}
m_{proton} = m_{u} + m_{u} + m_{d} = 938.25 \; MeV
\end{equation}
The $D_{4}-D_{5}-E_{6}$ model calculation is close to
the experimental value of 938.27 MeV.

\vspace{12pt}

The third generation fermion particles correspond to triples of
octonions.  There are $8^3$ = 512 such triples.

\vspace{12pt}

The triple $\{ 1,1,1 \}$ corresponds to the tau-neutrino.

\vspace{12pt}

The other 7 triples involving only $1$ and $e_{4}$ correspond
to the tauon:
$$\{ e_{4}, e_{4}, e_{4} \},
\{ e_{4}, e_{4}, 1 \},
\{ e_{4}, 1, e_{4} \},
\{ 1, e_{4}, e_{4} \},
\{ 1, 1, e_{4} \},
\{ 1, e_{4}, 1 \},
\{ e_{4}, 1, 1 \} $$,

The symmetry of the 7 tauon triples is the same as the
symmetry of the 3 down quarks, the 3 up quarks, and the electron,
so the tauon mass should be the same as the sum of the masses of
the first generation massive fermion particles.

\vspace{12pt}

Therefore the tauon mass 1.87704 GeV.

\vspace{12pt}

Note that all triples corresponding to the
tau and the tau-neutrino are colorless.

\vspace{12pt}

The beauty quark corresponds to 21 triples.
\newline
They are triples of the same form as the 7 tauon triples,
but for $1$ and $e_{3}$,  $1$ and $e_{5}$, and  $1$ and$ e_{7}$,
which correspond to the red, green, and blue beauty quarks,
respectively.

\vspace{12pt}

The seven triples of the red beauty quark correspond
to the seven triples of the tauon,
except that the beauty quark interacts with 6 $Spin(0,5)$
gravitons while the tauon interacts with only two.

\vspace{12pt}

The beauty quark constituent mass should be the tauon mass times the
third generation graviton factor 6/2 = 3, so the B-quark mass is
\newline
$m_{b}$ = 5.63111 GeV.

\vspace{12pt}

Note particularly that triples of the type $\{ 1, e_{3}, e_{5} \}$,
$\{ e_{3}, e_{5}, e_{7} \}$, etc.,
do not correspond to the beauty quark, but to the truth quark.

\vspace{12pt}

The truth quark corresponds to the remaining 483 triples, so the
constituent mass of the red truth quark is 161/7 = 23 times the
red beauty quark mass, and the red T-quark mass is
\begin{equation}
m_{t} = 129.5155 \; GeV
\end{equation}

The blue and green truth quarks are defined similarly.

\vspace{12pt}

The tree level T-quark constituent mass rounds off to 130 GeV.

\vspace{12pt}

These results when added up give a total mass of
\newline
third generation fermions:
\begin{equation}
\Sigma_{f_{3}} = 1,629 \; GeV
\end{equation}
\vspace{12pt}

\newpage

The second generation fermion calculations are:

The second generation fermion particles correspond
to pairs of octonions.
\newline
There are 82 = 64 such pairs.
\newline
The pair $\{ 1,1 \}$ corresponds to the $\mu$-neutrino.
\newline
The pairs $\{ 1, e_{4} \}$, $\{ e_{4}, 1 \}$, and
$\{ e_{4}, e_{4} \}$ correspond to the muon.
\newline
Compare the symmetries of the muon pairs to the symmetries
of the first generation fermion particles.
\newline
The pair $\{ e_{4}, e_{4} \}$ should correspond
to the $e_{4}$ electron.
\newline
The other two muon pairs have a symmetry group S2,
which is 1/3 the size of the color symmetry group S3
which gives the up and down quarks their mass of 312.75 MeV.

\vspace{12pt}

Therefore the mass of the muon should be the sum of
the $\{ e_{4}, e_{4} \}$ electron mass and
the $\{ 1, e_{4} \}$, $\{ e_{4}, 1 \}$ symmetry mass,
which is 1/3 of the up or down quark mass.

\vspace{12pt}

Therefore,  $m_{\mu}$ = 104.76 MeV.
\newline
\vspace{12pt}

Note that all pairs corresponding to
the muon and the $\mu$-neutrino are colorless.

\vspace{12pt}

The red, blue and green strange quark each corresponds
to the 3 pairs involving $1$ and $e_{3}$, $e_{5}$, or $e_{7}$.

\vspace{12pt}

The red strange quark is defined as the three pairs
$1$ and $e_{3}$, because $e_{3}$ is the red down quark.
\newline
Its mass should be the sum of two parts:
the $\{ e_{3}, e_{3} \}$ red down quark mass, 312.75 MeV, and
the product of the symmetry part of the muon mass, 104.25 MeV,
times the graviton factor.

\vspace{12pt}

Unlike the first generation situation,
massive second and third generation leptons can be taken,
by both of the colorless gravitons that
may carry electric charge, into massive particles.

\vspace{12pt}

Therefore the graviton factor for the second and
\newline
third generations is 6/2 = 3.

\vspace{12pt}

Therefore the symmetry part of the muon mass times
the graviton factor 3 is 312.75 MeV, and
the red strange quark constituent mass is
$$m_{s} = 312.75 \; MeV + 312.75 \; MeV = 625.5 \; MeV$$

The blue strange quarks correspond to the
three pairs involving $e_{5}$,
the green strange quarks correspond to the
three pairs involving $e_{7}$,
and their masses are determined similarly.

\vspace{12pt}

The charm quark corresponds to the other 51 pairs.
Therefore, the mass of the red charm quark should
be the sum of two parts:

\vspace{12pt}

the $\{ e_{1}, e_{1} \}$, red up quark mass, 312.75 MeV; and

\vspace{12pt}

the product of the symmetry part of the strange quark
mass, 312.75 MeV, and

\vspace{12pt}

the charm to strange octonion number factor 51/9,
which product is 1,772.25 MeV.

\vspace{12pt}

Therefore the red charm quark constituent mass is
$$m_{c} = 312.75 \; MeV + 1,772.25 \; MeV = 2.085 \; GeV$$

The blue and green charm quarks are defined similarly,
and their masses are calculated similarly.

\vspace{12pt}

These results when added up give a total mass of
second generation fermions:
\begin{equation}
\Sigma_{f_{2}} = 32.9 \; GeV
\end{equation}

\newpage

\vspace{12pt}

\subsubsection{Massless Neutrinos and Parity Violation}

\vspace{12pt}

It is required (as an ansatz or part of the
$D_{4}-D_{5}-E_{6}$ model)
that the charged $W_{\pm}$ neutrino-electron interchange
\newline
must be symmetric with the electron-neutrino interchange,
\newline
so that the absence of right-handed neutrino particles requires
\newline
that the charged $W_{\pm}$ $SU(2)$
weak bosons act only on left-handed electrons.

\vspace{12pt}

It is also required (as an ansatz or part of the
$D_{4}-D_{5}-E_{6}$ model) that each gauge boson must
act consistently on the entire Dirac fermion particle
sector, so that the charged $W_{\pm}$ $SU(2)$ weak bosons
act only on left-handed fermions of all types.

\vspace{12pt}

Therefore, for the charged $W_{\pm}$ $SU(2)$ weak bosons,
the 4-dimensional spinor fields $S_{8\pm}$ contain only
left-handed particles and right-handed antiparticles.
So, for the charged $W_{\pm}$ $SU(2)$ weak bosons,
$S_{8\pm}$ can be denoted $S_{8 \pm L}$.

\vspace{12pt}

The neutral $W_{0}$ weak bosons do not interchange Weyl
neutrinos with Dirac fermions, and so may not entirely
be restricted to left-handed spinor particle fields
$S_{8\pm L}$, but may have a component that acts on
the full right-handed and left-handed spinor particle
fields $S_{8\pm} = S_{8\pm L} + S_{8\pm R}$.

\vspace{12pt}

However, the neutral $W_{0}$ weak bosons are related to
the charged $W_{\pm}$ weak bosons by custodial $SU(2)$
symmetry, so that the left-handed component of the
neutral $W_{0}$ must be equal to the left-handed (entire)
component of the charged $W_{\pm}$.

\vspace{12pt}

Since the mass of the $W_{0}$ is greater than the mass
of the $W_{\pm}$, there remains for the $W_{0}$ a component
acting on the full $S_{8\pm} = S_{8\pm L} + S_{8\pm R}$
spinor particle fields.

\vspace{12pt}

Therefore the full $W_{0}$ neutral weak boson interaction
is proportional to
$(m_{W_{\pm}}^{2} / m_{W_{0}}^{2})$ acting on $S_{8\pm L}$
and
\newline
 $(1 - (m_{W_{\pm}}^{2} / m_{W_{0}}^{2}))$ acting
on $S_{8\pm} = S_{8\pm L} + S_{8\pm R}$.

\vspace{12pt}

If $(1 - (m_{W_{\pm}}2 / m_{W_{0}}^{2}))$ is defined to be
$\sin{\theta_{w}}^{2}$ and denoted by $\xi$, and

\vspace{12pt}

if the strength of the $W_{\pm}$ charged weak force
(and of the custodial $SU(2)$ symmetry) is denoted by $T$,

\vspace{12pt}

then the $W_{0}$ neutral weak interaction can be written as:

\vspace{12pt}

$W_{0L} \sim T + \xi$ and $W_{0R} \sim \xi$.

\vspace{12pt}

The $D_{4}-D_{5}-E_{6}$ model allows calculation of
the Weinberg angle $\theta_{w}$, by
\begin{equation}
m_{W_{+}} = m_{W_{-}} = m_{W_{0}} \cos{\theta_{w}}
\end{equation}

The Hopf fibration of $S^{3}$ as
$S^{1} \rightarrow  S^{3} \rightarrow  S^{2}$
gives a decomposition of the $W$ bosons
into the neutral $W_{0}$ corresponding to $S^{1}$ and
the charged pair $W_{+}$ and $W_{-}$ corresponding
to $S^{2}$.

\vspace{12pt}

The mass ratio of the sum of the masses of
$W_{+}$ and $W_{-}$ to
the mass of $W_{0}$
should be the volume ratio of
the $S^{2}$ in $S^{3}$ to
the $S^{1}$ in ${S3}$.

\vspace{12pt}

The unit sphere $S^{3} \subset R^{4}$ is
normalized by $1 / $2.

\vspace{12pt}

The unit sphere $S^{2} \subset R^{3}$ is
normalized by $1 / \sqrt{3}$.

\vspace{12pt}

The unit sphere $S^{1} \subset R^{2}$ is
normalized by $1 / \sqrt{2}$.

\vspace{12pt}

The ratio of the sum of the $W_{+}$ and $W_{-}$ masses to
the $W_{0}$ mass should then be
$(2  / \sqrt{3}) V(S^{2}) / (2 / \sqrt{2}) V(S^{1}) =
1.632993$.

\vspace{12pt}

The sum
$\Sigma_{m_{W}} = m_{W_{+}} + m_{W_{-}} + m_{W_{0}}$
has been calculated to be
$v \sqrt{\alpha_{w}} = 517.798  \sqrt{0.2534577} =
260.774 \; GeV$.

\vspace{12pt}

Therefore,
$\cos{\theta_{w}}^{2} = m_{W_{\pm}}^{2 } /
m_{W_{0}}^{2} = (1.632993/2)^{2} = 0.667$ , and

\vspace{12pt}

$\sin{\theta_{w}}^{2} = 0.333$,
so $m_{W_{+}} = m_{W_{-}} = 80.9 \; GeV$, and
$m_{W_{0}} = 98.9 \; GeV$.

\vspace{12pt}

\subsection{Corrections for $m_{Z}$ and $\theta_{w}$}

\vspace{12pt}

The above values must be corrected for the fact that
only part of the $w_{0}$ acts through the
parity violating $SU(2)$ weak force and the rest
acts through a parity conserving $U(1)$
electromagnetic type force.

\vspace{12pt}

In the $D_{4}-D_{5}-E_{6}$ model, the weak
parity conserving $U(1)$ electromagnetic type force
acts through the $U(1)$ subgroup of $SU(2)$,
which is not exactly like the $D_{4}-D_{5}-E_{6}$
electromagnetic $U(1)$ with force strength
$\alpha_{E} = 1 / 137.03608 = e^{2}$.

\vspace{12pt}

The $W_{0}$ mass $m_{W_{0}}$ has two parts:

\vspace{12pt}

the parity violating $SU(2)$ part $m_{W_{0\pm}}$ that is
equal to $m_{W_{\pm}}$ ; and

\vspace{12pt}

the parity conserving part $m_{W_{00}}$ that acts like a
heavy photon.

\vspace{12pt}

As $m_{W_{0}}$ = 98.9 GeV = $m_{W_{0\pm}} + m_{W_{00}}$, and
as $m_{W_{0\pm}} = m_{W_{\pm}} = 80.9 \; GeV$,
we have $m_{W_{00}} = 18 \; GeV$.

\vspace{12pt}

Denote by $\tilde{\alpha_{E}} = \tilde{e}^{2}$ the force
strength of the weak parity conserving $U(1)$
electromagnetic type force that acts through the
$U(1)$ subgroup of $SU(2)$.

\vspace{12pt}

The $D_{4}-D_{5}-E_{6}$ electromagnetic force strength
$\alpha_{E} = e^{2} = 1 / 137.03608$ was calculated using
the volume $V(S^{1})$ of an $S^{1} \subset R^{2}$,
normalized by $1 / \sqrt{2}$.

\vspace{12pt}

The $\tilde{\alpha_{E}}$ force is part of the $SU(2)$ weak
force whose strength $\alpha_{w} = w^{2}$ was calculated
using the volume $V(S^{2})$ of an $S^{2} \subset  R^{3}$,
normalized by $1  / \sqrt{3}$.

\vspace{12pt}

Also, the $D_{4}-D_{5}-E_{6}$ electromagnetic force
strength $\alpha_{E} = e^{2}$ was calculated using a
4-dimensional spacetime with global structure of
the 4-torus $T^{4}$ made up of four $S^{1}$ 1-spheres,

\vspace{12pt}

while the $SU(2)$ weak force strength
$\alpha_{w} = w^{2}$ was calculated using two 2-spheres
$S^{2} \times S^{2}$, each of which contains one 1-sphere of
the $\tilde{\alpha_{E}}$ force.

\vspace{12pt}

Therefore
$\tilde{\alpha_{E}} = \alpha_{E} (\sqrt{2} /
\sqrt{3})(2 / 4) = \alpha_{E} / \sqrt{6}$,
\newline
 $\tilde{e}  = e / 4 \sqrt{6} = e / 1.565$ , and

\vspace{12pt}

the mass $m_{W_{00}}$ must be reduced to an effective value

\vspace{12pt}

$m_{W_{00}eff} = m_{W_{00}} / 1.565$ = 18/1.565 = 11.5 GeV

\vspace{12pt}

for the $\tilde{\alpha_{E}}$ force to act like
an electromagnetic force in the 4-dimensional
spacetime of the $D_{4}-D_{5}-E_{6}$ model:

\vspace{12pt}

$\tilde{e} m_{W_{00}} = e (1/5.65) m_{W_{00}} = e m_{Z_{0}}$,

\vspace{12pt}

where the physical effective neutral weak boson is
denoted by $Z$ rather than $W_{0}$.

\vspace{12pt}

Therefore, the correct $D_{4}-D_{5}-E_{6}$ values for
weak boson masses and the Weinberg angle are:

\vspace{12pt}

$m_{W_{+}} = m_{W_{-}} = 80.9 \; GeV$;

\vspace{12pt}

$m_{Z} = 80.9 +11.5 = 92.4 \; GeV$; and

\vspace{12pt}

$\sin{\theta_{w}}^{2} = 1 - (m_{W_{\pm}} /
m_{Z})^{2} = 1 - 6544.81/8537.76 = 0.233$.

\vspace{12pt}

Radiative corrections are not taken into account here,
and may change the $D_{4}-D_{5}-E_{6}$ value somewhat.

\vspace{12pt}

\newpage

\subsection{K-M Parameters.}

The following formulas use the above masses to
calculate Kobayashi-Maskawa parameters:

\begin{equation}
phase \; angle \; \epsilon = \pi / 2
\end{equation}

\begin{equation}
\sin{\alpha} = [m_{e}+3m_{d}+3m_{u}] /
\sqrt{ [m_{e}^{2}+3m_{d}^{2}+3m_{u}^{2}] +
[m_{\mu}^{2}+3m_{s}^{2}+3m_{c}^{2}] }
\end{equation}

\begin{equation}
\sin{\beta} = [m_{e}+3m_{d}+3m_{u}] /
\sqrt{ [m_{e}^{2}+3m_{d}^{2}+3m_{u}^{2}] +
[m_{\tau}^{2}+3m_{b}^{2}+3m_{t}^{2}] }
\end{equation}

\begin{equation}
\sin{\tilde{\gamma}} = [m_{\mu}+3m_{s}+3m_{c}] /
\sqrt{ [m_{\tau}^{2}+3m_{b}^{2}+3m_{t}^{2}] +
[m_{\mu}^{2}+3m_{s}^{2}+3m_{c}^{2}] }
\end{equation}

\begin{equation}
\sin{\gamma} = \sin{\tilde{\gamma}}
\sqrt{\Sigma_{f_{2}} / \Sigma_{f_{1}}}
\end{equation}

\vspace{12pt}

The resulting Kobayashi-Maskawa parameters are:

\begin{equation}
\begin{array}{|c|c|c|c|}
\hline
& d & s & b
  \\
\hline
u & 0.975 & 0.222 & -0.00461 i  \\
c & -0.222 -0.000191 i & 0.974 -0.0000434 i & 0.0423  \\
t & 0.00941 -0.00449 i & -0.0413 -0.00102 i & 0.999  \\
\hline
\end{array}
\end{equation}

\end{document}